\documentclass[10pt,journal,compsoc]{IEEEtran}
\ifCLASSOPTIONcompsoc
\usepackage[nocompress]{cite}
\else
\usepackage{cite}
\fi
\usepackage{algorithm}
\usepackage{algorithmic}
\usepackage{xcolor}

\usepackage{url}
\usepackage{graphicx}
\usepackage{amsthm}
\usepackage{amssymb}
\usepackage{amsfonts}
\usepackage{amsmath}
\usepackage{multirow}
\newcommand{\eat}[1]{}
\newcommand{\etal}{{et al.~}}       
\newcommand{\ie}{{i.e.~}}           
\newcommand{\etc}{{etc.~}}         
\newcommand{\wrt}{{w.r.t.~}}         
\newcommand{\aka}{{a.k.a.~}}        

\newtheorem{definition}{Definition}
\newtheorem{problem}{Problem}
\newtheorem{example}{Example}

\newtheorem{corollary}{Corollary}

\newtheorem{theorem}{Theorem}
\makeatletter
\newcommand*{\indep}{%
	\mathbin{%
		\mathpalette{\@indep}{}%
	}%
}
\newcommand*{\nindep}{%
	\mathbin{
		\mathpalette{\@indep}{\not}
	}%
}
\newcommand*{\@indep}[2]{%
	\sbox0{$#1\perp\m@th$}
	\sbox2{$#1=$}
	\sbox4{$#1\vcenter{}$}
	\rlap{\copy0}
	\dimen@=\dimexpr\ht2-\ht4-.2pt\relax
	\kern\dimen@
	{#2}%
	\kern\dimen@
	\copy0 
}

\def\astrightarrow{\put(0.2,-2.2){*}\rightarrow}
\def\astleftarrow{\leftarrow\put(-5,-2.2){*}}

\ifCLASSINFOpdf
\else
\fi
\begin{document}
		%
		\title{Local search for efficient causal effect estimation}
		%
		%
		%
		%
		
		\author{Debo~Cheng, Jiuyong~Li, Lin~Liu, Jiji~Zhang, Jixue~Liu
			and~Thuc~Duy~Le 
			\IEEEcompsocitemizethanks{\IEEEcompsocthanksitem D. Cheng, J. Li, L. Liu, J. Liu and T. Le are with UniSA STEM, University of South Australia, Adelaide,
				SA, 5095. E-mail: \{debo.cheng,jiuyong.li,liu.liu,jixue.liu,thuc.le\}@unisa.edu.au \protect\\
				\protect\\
				\IEEEcompsocthanksitem J. Zhang is with the Department of Religion and Philosophy, Hong Kong Baptist University, Hong Kong, China.  E-mail: zhangjiji@hkbu.edu.hk 
			}
			}
		
		%
		%

	\markboth{Preprint}%
	{Shell \MakeLowercase{\textit{et al.}}: Bare Demo of IEEEtran.cls for Computer Society Journals}
	%



	\IEEEtitleabstractindextext{%
		\begin{abstract}
			Causal effect estimation from observational data is a challenging problem, especially with high dimensional data and in the presence of unobserved variables.  
			The available data-driven methods for tackling the problem either provide an estimation of the bounds of a causal effect (\ie nonunique estimation) or have low efficiency. The major hurdle for achieving high efficiency while trying to obtain unique and unbiased causal effect estimation is how to find a proper adjustment set for confounding control in a fast way, given the huge covariate space and considering unobserved variables. In this paper, we approach the problem as a local search task for finding valid adjustment sets in data. We establish the theorems to support the local search for adjustment sets, and we show that unique and unbiased estimation can be achieved from observational data even when there exist unobserved variables. We then propose a data-driven algorithm that is fast and consistent under mild assumptions. We also make use of a frequent pattern mining method to further speed up the search of minimal adjustment sets for causal effect estimation. Experiments conducted on extensive synthetic and real-world datasets demonstrate that the proposed algorithm outperforms the state-of-the-art criteria/estimators in both accuracy and time-efficiency.
		\end{abstract}
		
		}

	\maketitle

	\IEEEdisplaynontitleabstractindextext
	
	%
	\IEEEpeerreviewmaketitle
	
\IEEEraisesectionheading{\section{Introduction}\label{sec:intro}}
\IEEEPARstart{O}{ne} of the fundamental tasks of causal inference is to estimate the causal effect (\aka treatment effect) of a treatment on the outcome. Causal effect estimation has attracted increasing attention in many fields such as computer sciences~\cite{pearl2009causality}, epidemiology~\cite{hernan2020causal}, psychology~\cite{spirtes2000causation} and econometrics~\cite{imbens2015causal}, \etc \emph{Randomised Controlled Trials} (RCTs) are regarded as the most reliable means to estimate causal effects~\cite{deaton2018understanding}, but RCTs are often infeasible due to ethical concerns and/or high expenses. Hence, estimating causal effects using observational data has been explored as an important alternative to RCTs.

With observational data, covariate adjustment is the main approach to eliminating confounding bias in causal effect estimation. \emph{Graphical causal modelling}~\cite{pearl2009causality,shpitser2012validity,van2019separators} provides a theoretical framework for determining a valid adjustment set. For example, given a causal DAG (directed acyclic graph) which represents the underlying causal mechanism, the \emph{back-door criterion}~\cite{pearl2009causality} can be employed to determine a valid adjustment set.

In many real-world applications, users do not know the underlying causal DAGs, so data-driven methods have to be used to estimate causal effects directly from observational data. However, with a data-driven approach, a unique estimation of a causal effect cannot be obtained without making certain assumptions. Specifically, from  observational data, generally, we can only learn a Markov equivalence class of causal structures with the causal graphical modelling approach~\cite{maathuis2015generalized,perkovic2018complete}, instead of a unique causal structure~\cite{spirtes2000causation,maathuis2015generalized}, and we do not know which structure in the Markov equivalence class is the true causal structure. Consequently, the causal effect estimation based on the Markov equivalence class of structures is a bound estimation, i.e. a set of all possible causal effects, each corresponding to a structure in the Markov equivalence class. For example, \cite{maathuis2009estimating} proposed IDA (Intervention when the DAG is Absent) to estimate causal effects from data satisfying causal sufficiency. The output of IDA is a bound estimation, as a result of the aforementioned non-uniqueness in structure learning. A bound estimation may have a wide range and thus may not give users a good indicator where the causal effect value stands. For example, users even may not know whether the causal effect is positive or negative. Therefore, it is desirable to obtain unique causal effect estimation.

Moreover, in practice, there usually exist unobserved (or latent) confounding variables, so the causal sufficiency assumption is not satisfied. A few methods have been developed for causal effect estimation from data with latent variables. For example, CE-SAT~\cite{hyttinen2015calculus} uses SAT-based inference to estimate causal effects from data with latent variables, but it can only handle a small number of variables (at most around 20). LVIDA (Latent Variable IDA)~\cite{malinsky2017estimating} employs FCI (fast causal inference~\cite{spirtes2000causation}) to search for a Markov equivalence class of maximal ancestral graphs (MAGs) and runs an IDA-like procedure on the equivalence class. Like IDA, it generally outputs a bound estimation, and is difficult to be scaled to large datasets with dozens of variables or more due to the high computational cost of global structure learning methods~\cite{chickering2004large,aliferis2010local}. The DICE algorithm~\cite{cheng2020causal} searches for adjustment sets in local causal structures, which significantly improves the efficiency, but DICE finds a superset of one or multiple adjustment sets and thus still provides only a bound estimation. The EHS algorithm~\cite{entner2013data} employs conditional independence tests to identify an adjustment set. It can return valid adjustment sets and unique causal effect estimations. However, EHS is very inefficient since it conducts an exhaustive search over all the combinations of variables for the conditional independence tests. Moreover, the adjustment sets identified by EHS can be large, containing redundant variables, which can result in inaccurate estimations of causal effects.

In this paper, we aim to tackle the two challenges (uncertainty in finding causal structures or adjustment sets from data and low efficiency) in data-driven causal effect estimation simultaneously. As discussed above, the existing methods have a high time complexity. For example, the time complexities of LVIDA and EHS are $O(p2^{p-1})$ in the worst case~\cite{colombo2012learning,entner2013data,malinsky2017estimating}, where $p$ denotes the number of variables. The advancement in local causal discovery techniques has led to significant improvement in the efficiency of causal structure learning from data~\cite{aliferis2010local}, but no research has been done to use local causal discovery methods to determine adjustment sets from data with latent variables. In this paper, we first develop a theorem to support the use of local search methods for determining proper adjustment sets from data with latent variables.  Then, we develop a new data-driven method, $CEELS$ (Causal Effect Estimation by Local Search) to find \emph{minimal} adjustment sets for \emph{unique} and \emph{unbiased} causal effect estimation from data with latent variables, under a testable assumption called the COSO (Cause of Or Sharing a latent confounder\footnote{A variable which shares a latent confounder with the treatment is known as a ``spouse'' of the treatment in standard causal modelling~\cite{richardson2002ancestral}.}  with the treatment Only) variable assumption (see Definition~\ref{def:COSO}). $CEELS$ is very fast as it conducts a local search and an efficient \emph{pattern mining strategy}~\cite{agrawal1994fast} for discovering adjustment sets.

In sum, the paper makes the following contributions:
\begin{itemize}
\item We establish several theoretical conclusions under mild assumptions, with which adjustment sets can be found by local search given observational data with latent variables. The use of local search reduces the search space of proper adjustment sets into $O(2^{q})$ where $q$ is the number of nodes adjacent to the outcome in a PAG, and $q\ll p$.
\item Based on the theoretical results, we develop an efficient data-driven algorithm, $CEELS$, for identifying valid adjustment sets from observational data with latent variables. To our best knowledge, $CEELS$ is the first practical data-driven method for finding valid adjustment sets from data with latent variables by local search. Experiments show that $CEELS$ is significantly more efficient than the algorithms based on global search.
\end{itemize}

\section{Preliminaries and background}
\label{sec:pre}
\subsection{Notation and Definitions}
\label{subsec:not}
Let $\mathcal{G}=(\mathbf{V}, \mathbf{E})$ be a graph with a set of nodes $\mathbf{V}=\{V_{1}, \dots, V_{p}\}$ (denoting random variables) and a set of edges $\mathbf{E} \subseteq \mathbf{V} \times \mathbf{V}$ (denoting the relationships between the nodes).

A \emph{directed graph} contains only directed edges ($\rightarrow$). A mixed graph may contain both directed and bi-directed edges ($\leftrightarrow$)~\cite{zhang2008causal,perkovic2018complete}. In a directed or mixed graph, two nodes are adjacent if there is an edge between them, and $V_i$ is a parent of $V_j$ (and $V_j$ is a child of $V_i$) if $V_i \rightarrow V_j$ appears in the graph. We use $Adj(V)$, $Pa(V)$ and $Ch(V)$ to denote the sets of all adjacent nodes, parents and children of $V$, respectively. In a mixed graph, if there is $V_{i}\leftrightarrow V_{j}$, $V_i$ and $V_j$ share a latent confounder that affects both $V_i$ and $V_j$. $Sl(V)$ denotes the set of all nodes that share a latent confounder with $V$. A directed or mixed graph can represent causal relationships. In the context of this paper, in a mixed graph, a visible edge $V_i\rightarrow V_j$ (edge visibility is defined in Definition~\ref{Visibility}) indicates that $V_i$ is a direct cause of $V_j$, and $V_i \leftrightarrow V_j$ indicates that $V_i$ and $V_j$ are not each other's direct causes and there is a latent common cause for $V_i$ and $V_j$.

In a mixed graph $\mathcal{G}$, a path $\pi$ from $V_{1}$ to $V_{p}$ comprises a sequence of nodes $\langle V_{1}, \dots, V_{p}\rangle$ with every pair of successive nodes being adjacent. A path $\pi$ is a directed path if all edges along it are directed, i.e. $V_{1} \rightarrow\ldots \rightarrow V_{p}$. A directed path is also called a causal path. In a directed path $\pi$, $V_i$ is an ancestor of $V_j$ and $V_j$ is a descendant of $V_i$ if all arrows along $\pi$ point to $V_j$. The sets of ancestors and descendants of $V_i$ are denoted as $An(V_i)$ and $De(V_i)$, respectively.

\begin{figure}[t]
\centering
\includegraphics[scale=0.23]{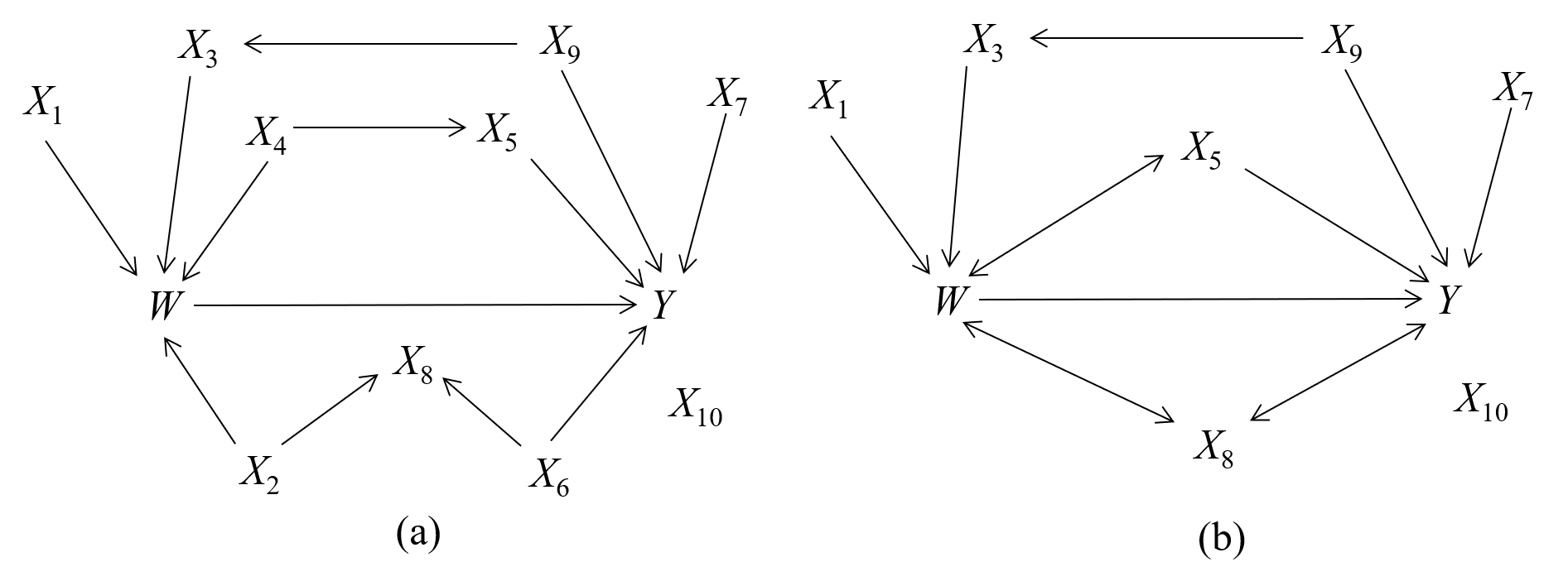}
\caption{An example used to explain various terms used in graphical causal modelling. (a) is a DAG for generating synthetic datasets in~\cite{witte2019covariate} and (b) is its corresponding MAG when $X_2, X_4, X_6$ are unobserved.}
\label{fig:causaldiagram}
\end{figure}

From data, one can identify a Markov equivalence class of the underlying causal graph~\cite{spirtes2000causation,pearl2009causality}. Partial mixed graph is used to represent the Markov equivalence class. There are three types of end marks for edges in a partial mixed graph $\mathcal{G}$: arrowhead $(<)$, tail $(-)$, and circle $(\circ)$ (denoting that the orientation of the edge is uncertain)~\cite{zhang2008completeness}. Let $\ast$ be an arbitrary edge mark. $V_{i}$ is a collider on a path $\pi$ if $V_{i-1}\astrightarrow V_{i} \astleftarrow V_{i+1}$ is in $\mathcal{G}$. A collider path is a path with every non-endpoint node being a collider. For example, in Fig.~\ref{fig:causaldiagram} (a) or (b), $X_8$ is a collider. A path of length one is a trivial collider path. $V_j$ is a \emph{definite non-collider} on $\pi$ if there exists at least an edge out of $V_j$ on $\pi$, or both edges have a circle mark at $V_j$ and the subpath $\langle V_i, V_j, V_k\rangle$ is an unshielded triple (i.e. $V_i$ and $V_k$ are not adjacent)~\cite{zhang2008causal}. A node is of a \emph{definite status} on a path if it is a collider or a definite non-collider on the path. A path $\pi$ is in \emph{definite status} if every non-endpoint node on $\pi$ is in definite status~\cite{perkovic2018complete}.

A directed cycle forms in a directed or mixed graph if there exists $V_i \rightarrow V_j$ and $V_j\in An(V_i)$. A directed acyclic graph (DAG) is a directed graph containing no directed cycles. In a mixed graph, an almost directed cycle occurs when $V_i\leftrightarrow V_j$ is in the mixed graph and $V_j\in An(V_i)$. An almost directed cycle is defined (and named) based on the fact that removing the arrowhead at $V_i$ in $V_i\leftrightarrow V_j$ results in a directed cycle~\cite{zhang2008causal}.

\begin{definition}[Markov property~\cite{pearl2009causality}]
\label{Markov condition}
Given a DAG $\mathcal{G}=(\mathbf{V}, \mathbf{E})$ and the joint probability distribution of $\mathbf{V}$ $(P(\mathbf{V}))$, $\mathcal{G}$ satisfies the Markov property if for $\forall V_i \in \mathbf{V}$, $V_i$ is probabilistically independent of all of its non-descendants given $Pa(V_i)$.
\end{definition}

\begin{definition}[Faithfulness~\cite{spirtes2000causation}]
\label{Faithfulness}
A DAG $\mathcal{G}=(\mathbf{V}, \mathbf{E})$ is faithful to a joint distribution $P(\mathbf{V})$ over the set of variables $\mathbf{V}$ if and only if every independence present in  $P(\mathbf{V})$ is entailed by $\mathcal{G}$ and satisfies the Markov property. A joint distribution $P(\mathbf{V})$ over the set of variables $\mathbf{V}$ is faithful to the DAG $\mathcal{G}$ if and only if the DAG $\mathcal{G}$ is faithful to the joint distribution $P(\mathbf{V})$.
\end{definition}

When the faithfulness assumption is satisfied between a probability distribution and a DAG of a set of variables, the dependency/independency relations among the variables can be read from the DAG.

\begin{definition}[Causal sufficiency~\cite{spirtes2000causation}]
\label{def:causuf}
A given dataset satisfies causal sufficiency if for every pair of observed variables, all their common causes are observed.
\end{definition}

In this work, ancestral graphs are used to represent the data generating process that involves latent variables~\cite{richardson2002ancestral}.
\begin{definition}[Ancestral graph]
\label{Ancestral graph}
An ancestral graph is a mixed graph that does not contain directed cycles or almost directed cycles.
\end{definition}

The criterion of m-separation is a natural extension of the well-known d-separation criterion~\cite{pearl2009causality} to mixed graphs.
\begin{definition}[m-separation~\cite{richardson2002ancestral,richardson2003causal}]
\label{m-separation}
In a mixed graph $\mathcal{M}=(\mathbf{V}, \mathbf{E})$, a path $\pi$ between $V_{i}$ and $V_{j}$ is said to be m-separated by a set of nodes $\mathbf{Z}\subseteq \mathbf{V}\setminus\{V_i, V_j\}$ (possibly $\emptyset$) if $\pi$ contains a subpath $\langle V_l, V_k, V_s\rangle$ such that the middle node $V_k$ is a non-collider on $\pi$ and $V_k \in \mathbf{Z}$; or $\pi$ contains $V_l\astrightarrow V_k\astleftarrow V_s$ such that $V_k \notin \mathbf{Z}$ and no descendant of $V_k$ is in $\mathbf{Z}$. 
Two nodes $V_{i}$ and $V_{j}$ are said to be m-separated by $\mathbf{Z}$ in $\mathcal{M}$ if every path between $V_{i}$ and $V_{j}$ are m-separated by $\mathbf{Z}$; otherwise, they are said to be m-connected by $\mathbf{Z}$ in $\mathcal{M}$.
\end{definition}
\begin{definition}[Maximal ancestral graph (MAG)]
\label{MAG}
An ancestral graph $\mathcal{M}=(\mathbf{V}, \mathbf{E})$ is a MAG when every pair of non-adjacent nodes $V_{i}$ and $V_{j}$ in $\mathcal{M}$ is m-separated by a set $\mathbf{Z}\subseteq \mathbf{V}\backslash \{V_{i}, V_{j}\}$.
\end{definition}

Note that all nodes in a MAG are in definite status~\cite{van2014constructing}. Finally, the following is needed for our discussion.

\begin{definition}[Visibility~\cite{zhang2008causal}]
\label{Visibility}
Given a MAG $\mathcal{M}=(\mathbf{V}, \mathbf{E})$, a directed edge $V_{i}\rightarrow V_{j}$ is visible if there is a node $V_{k}\notin Adj(V_{j})$, such that either there is an edge between $V_{k}$ and $V_{i}$ that is into $V_{i}$, or there is a collider path between $V_{k}$ and $V_{i}$ that is into $V_{i}$ and every node on this path is a parent of $V_{j}$. Otherwise, $V_{i}\rightarrow V_{j}$ is said to be invisible.
\end{definition}

The edge $V_i\rightarrow V_j$ in a MAG is visible means that there are no latent confounders between $V_i$ and $V_j$. Otherwise, there exists a latent confounder between $V_i$ and $V_j$.

\subsection{Causal Effect Estimation by Adjustment}
\label{subsec:causal}
Let $W$ be a binary variable indicating treatment status ($w = 1$ for being treated and $w=0$ for control), and $Y$ the outcome of interest, and $\mathbf{X}$ a set of pretreatment variables, i.e. none of the descendant nodes of $W$ or $Y$ is in $\mathbf{X}$. The pretreatment assumption is a realistic assumption as it reflects how a sample is obtained in many application areas such as economics and epidemiology~\cite{imbens2015causal,hill2011bayesian,wager2018estimation}, and the assumption is commonly used by existing methods~\cite{vanderweele2011new,de2011covariate,entner2013data}. We further make the common assumptions of Markov property and faithfulness \wrt the distribution of data and the underlying data generating process represented by a causal DAG~\cite{spirtes2000causation}. Under both assumptions, it is possible to discover the set of adjacent nodes of a specific variable from data by using conditional independence tests~\cite{aliferis2010local,maathuis2015generalized}, and with the pretreatment assumption. 

The average causal effect of $W$ on $Y$, i.e. $ACE(W, Y)$, is defined as:
\begin{equation}
ACE(W, Y)=\mathbf{E}(Y\mid do(w=1))-\mathbf{E}(Y\mid do(w=0))
\end{equation}

\noindent where $do()$ denotes the do-operator indicating that the value of $W$ is set to a specific value $w$.  With a valid adjustment set $\mathbf{Z}$, $ACE(W, Y)$ can be obtained unbiasedly from data as follows.
\begin{equation}\label{eq002}
\begin{split}
	ACE(W, Y)&=  \Sigma_{\mathbf{z}}[\mathbf{E}(Y\mid w=1,\mathbf{Z}=\mathbf{z})- \\ & \mathbf{E}(Y\mid w=0,\mathbf{Z}=\mathbf{z})]P(\mathbf{Z}=\mathbf{z})
\end{split}
\end{equation}

When the underlying causal DAG is known, the \emph{back-door criterion}~\cite{pearl2009causality} can be used to identify an adjustment set $\mathbf{Z}$ from the known causal DAG, if such a set exists. The back-door criterion is a well-known criterion for determining an adjustment set in a given DAG. Assuming that we have a DAG $\mathcal{G}=(\mathbf{V, E})$ where $\mathbf{V} =\mathbf{X}\cup\{W, Y\}$ and a dataset data on $\mathbf{V}$. The back-door criterion can be used directly to find an adjustment set $\mathbf{Z}\subseteq \mathbf{X}$ in the given $\mathcal{G}$.

\begin{definition}[Back-door criterion~\cite{pearl1995causal}]
\label{def:backdoorcrite}
In a DAG $\mathcal{G}=(\mathbf{V, E})$, where $\mathbf{V}=\{W, Y\}\cup \mathbf{X}$ and $(W, Y)$ is an ordered pair of variables. A set of variables $\mathbf{Z}\subseteq \mathbf{X}$ is said to satisfy the back-door criterion in the given DAG $\mathcal{G}$ if
\begin{enumerate}
	\item $\mathbf{Z}$ does not contain a descendant node of $W$; and
	\item $\mathbf{Z}$ blocks every back-door path between $W$ and $Y$ (i.e. any path between $W$ and $Y$ containing an arrow into $W$).
\end{enumerate}	
A set $\mathbf{Z}$ is called a \emph{back-door set} relative to $(W, Y)$ in $\mathcal{G}$ if $\mathbf{Z}$ satisfies the criterion relative to  $(W, Y)$ in $\mathcal{G}$.
\end{definition}

For example, in the DAG in Fig.~\ref{fig:causaldiagram}(a), $\{X_3, X_4\}$ satisfies the back-door criterion relative to the ordered pair of variables $(W, Y)$. However, causal sufficiency is often violated in real applications since not all variables are observed. In this case, a MAG is employed to represent a system that involves latent variables, Maathuis and Colombo~\cite{maathuis2015generalized} generalised the concept of back-door path in a DAG to the concept of generalised back-door path in a MAG \footnote{Note that the definitions of generalised back-door path, amenability, forbidden set and generalised adjustment criterion are all introduced regarding a single treatment variable and a single outcome variable, while the original definitions are discussed regarding a set of treatment variables  $\mathbf{W}$ and a set of outcome variables $\mathbf{Y}$.} 

\begin{definition}[Generalised back-door path~\cite{maathuis2015generalized}]
\label{Def:genbackdoorpath}
In a given MAG $\mathcal{M} = (\mathbf{V, E})$, where $\mathbf{V}=\{W, Y\}\cup \mathbf{X}$ and $(W, Y)$ is an ordered pair of variables.  A path between $W$ and $Y$ in $\mathcal{M}$ is a generalised back-door path from $W$ to $Y$ if it does not have a visible edge out of $W$.
\end{definition} 

To determine an adjustment set \wrt the ordered pair of variables $(W, Y)$ in a given MAG, Perkovi{\'c} \etal\cite{perkovic2018complete} extended the back-door criterion and developed the following Generalised Adjustment Criterion (GAC). To introduce GAC, first some additional definitions are provided.

\begin{definition}[Amenability~\cite{van2014constructing}]
\label{Amenability}
Given a MAG $\mathcal{M} = (\mathbf{V, E})$, where $\mathbf{V}=\{W, Y\}\cup \mathbf{X}$ and $(W, Y)$ is an ordered pair of variables. $\mathcal{M}$ is adjustment amenable w.r.t. $(W, Y)$ if each directed path from $W$ to $Y$ in $\mathcal{M}$ starts with a visible edge.
\end{definition}

The forbidden set w.r.t. the ordered pair of variables $(W, Y)$ in a MAG $\mathcal{M}$ is defined as below.

\begin{definition}[Forbidden set; $Forb(W, Y, \mathcal{M})$~\cite{perkovic2018complete}]
\label{def:forbiden}
Given a MAG $\mathcal{M}=(\mathbf{V}, \mathbf{E})$ where $\mathbf{V}=\{W, Y\}\cup \mathbf{X}$ and $(W, Y)$ is an ordered pair of variables. The forbidden set w.r.t. $(W, Y)$ is
$Forb(W, Y, \mathcal{M})=\{X\in \mathbf{V}: X\in De(W)$, $X$ lies on a causal path from $W$ to $Y$ in $\mathcal{M}\}$.
\end{definition}
In this work, $Forb(W, Y, \mathcal{M}) = \emptyset$ since $\mathbf{X}$ contains pretreatment variables only. 

\begin{definition}[Generalised Adjustment Criterion (GAC) for a MAG~\cite{perkovic2018complete,van2014constructing}]
\label{def:GAC}
Given a MAG $\mathcal{M}=(\mathbf{V}, \mathbf{E})$, where $\mathbf{V}=\{W, Y\}\cup \mathbf{X}$ and $(W, Y)$ is an ordered pair of variables. A set of nodes $\mathbf{Z}\subseteq \mathbf{X}$ satisfies the GAC relative to $(W, Y)$, i.e. $\mathbf{Z}$ is an adjustment set if (i) $\mathcal{M}$ is adjustment amenable relative to $(W, Y)$, and (ii) $\mathbf{Z}\cap Forb(W, Y, \mathcal{M})= \emptyset$, and (iii) all definite statues non-causal paths between $W$ and $Y$ are blocked (i.e. m-separated) by $\mathbf{Z}$.
\end{definition}

In the MAG shown in Fig.~\ref{fig:causaldiagram}(b), $\{X_3, X_5\}$ satisfies the GAC criterion relative to the ordered pair of variables $(W, Y)$. Similar to the back-door criterion, which requires a DAG to be known, GAC requires a MAG to be given. The EHS condition~\cite{entner2013data} is able to infer an adjustment set directly from data based on conditional independence tests when assuming the Markov property and faithfulness. 
\begin{theorem}[EHS condition~\cite{entner2013data}]
\label{Theo:EHSsearch}
Given a dataset $\mathbf{D}$ with the treatment $W$, outcome $Y$ and a set of pretreatment variables $\mathbf{X}$. If there exists a variable $S\in\mathbf{X}$ and a set $\mathbf{Z}\subseteq\mathbf{X}\setminus \{S\}$ such that $S\nindep Y\mid\mathbf{Z}$ and $S\indep Y\mid \mathbf{Z}\cup\{W\}$, then $\mathbf{Z}$ is an adjustment set.
\end{theorem}
The EHS condition supports a data-driven approach to determining an adjustment set directly from data with latent variables. However, as mentioned previously, the EHS algorithm~\cite{entner2013data} is inefficient (with time complexity of $\mathbf{O}(2^{|\mathbf{X}|})$), as it requires a global search for discovering the adjustment sets. A causal graph is sufficient for determining whether a causal effect is identifiable. However, the complexly of discovering a complete causal graph in data is super exponential to the number of variables~\cite{chickering2004large,colombo2012learning}, and hence it may not be possible to learn a causal graph from a dataset, or if using a heuristic algorithm, the causal graph learned may not be reliable. In contrast, the complexity for finding a local causal structure is polynomial to the number of variables~\cite{aliferis2010local}. Therefore a local structure can be learned from a dataset faster and more reliably, and the identification of adjustment sets in a local stricture is faster and more reliable too. In this paper, we develop new theorems to support a local search for adjustment set discovery based on the EHS condition and propose an efficient algorithm based on the developed theorems.

\section{Local search for efficient adjustment set discovery}
\label{sec:ce2ls}
\subsection{Theorems for Local Search of Adjustment Sets}
\label{subsec:theo}
We now define the problem to be solved in the paper, and then present the theorems for supporting adjustment set discovery by local search.

\begin{problem}
Given a dataset $\mathbf{D}$ generated from an underlying MAG $\mathcal{M}=(\mathbf{V}, \mathbf{E})$, where $\mathbf{V}=\{W, Y\}\cup \mathbf{X}$, and $W\rightarrow Y$ is in $\mathcal{M}$ as a visible edge.
We aim to identify a valid adjustment set $\mathbf{Z} \subseteq \mathbf{X}$ \wrt $(W, Y)$ in the local structures around $Y$ and $W$, i.e. $Adj(Y)\setminus\{W\}$ or $Adj(W)\cup Adj(Y)\setminus\{W, Y\}$ based on the conditional independence tests in the EHS condition (Theorem \ref{Theo:EHSsearch}).
\end{problem}

Note that we do not assume that the underlying MAG is known. In our problem setting (i.e. data generation is governed by an underlying MAG containing $W\rightarrow Y$ as a visible edge and the pretreatment assumption holds), based on GAC (Definition~\ref{def:GAC}), we have the following corollary.
\begin{corollary}
\label{coro:problemsetting}
In the underlying MAG $\mathcal{M}=(\mathbf{V}, \mathbf{E})$, where $\mathbf{V}=\{W, Y\}\cup \mathbf{X}$ and $(W, Y)$ is an ordered pair of variables. If $\exists\mathbf{Z}\subseteq \mathbf{X}$ which blocks all non-causal paths from $W$ to $Y$, then $\mathbf{Z}$ is a valid adjustment set, i.e. $ACE(W, Y)$ can be estimated unbiasedly by adjusting for $\mathbf{Z}$ as in Eq.(\ref{eq002}).
\end{corollary}
\begin{proof}
Under the pretreatment assumption, $W\rightarrow Y$ is the only directed path from $W$ to $Y$ in $\mathcal{M}$, and $W\rightarrow Y$ is a visible edge, so $\mathcal{M}$ is adjustment amenable \wrt $(W, Y)$ based on Definition~\ref{Amenability}. Moreover, $Forb(W, Y, \mathcal{M})$ is $\emptyset$ because for $\forall X\in \mathbf{X}$, $X\notin De(W)$ in $\mathcal{M}$. Hence, the first two conditions of GAC are satisfied. There exists a set $\mathbf{Z}\subseteq \mathbf{X}$ that blocks all non-causal paths from $W$ to $Y$, then the third condition of GAC is also satisfied, and $\mathbf{Z}$ is an adjustment set \wrt $(W, Y)$. Therefore, the causal effect of $W$ on $Y$ can be estimated unbiasedly by adjusting for $\mathbf{Z}$ as in Eq.(2).
\end{proof}

One way to identify such adjustment sets is to search for a set that satisfies the EHS condition. The question we are going to answer is that under what conditions we can identify them by \textit{local} search. The following theorem showing that if there is a variable $Q$ that is adjacent to $W$ but not to $Y$, then the local search is guaranteed to find a set that satisfies the EHS condition.

\begin{theorem}
\label{coro:corCIT}
Suppose that there exists a cause $Q$ of $W$ such that $Q\in Adj(W)\setminus Adj(Y)$. Then there exists a set $\mathbf{Z}\subseteq Adj(Y) \setminus \{W\}$ that satisfies the EHS condition with $Q$, namely, $Q\nindep Y \mid\mathbf{Z}$ and $Q\indep Y\mid\mathbf{Z}\cup \{W\}$, and so $\mathbf{Z}$ is a valid adjustment set.
\end{theorem}
\begin{proof}
Firstly, we prove that $Pa(Y)\setminus \{W\}$ satisfies GAC. From the proof of Corollary~\ref{coro:problemsetting}, the first two conditions of GAC are satisfied in our problem setting. In a MAG $\mathcal{M}$, all nodes are in definite status such that all paths in the MAG are in definite status. In our problem setting, $\mathbf{X}$ includes the pretreatment variables only and there exists a visible edge $W\rightarrow Y$ in $\mathcal{M}$, thus all paths from $W$ to $Y$ (excluding the path $W\rightarrow Y$) in $\mathcal{M}$ are non-causal paths~\cite{zhang2008causal,perkovic2018complete}. Therefore, according to Definition~\ref{Def:genbackdoorpath}, a path from $W$ to $Y$ except $W\rightarrow Y$ is a generalised back-door path from $W$ to $Y$. So in the following, we prove that $Pa(Y)\setminus \{W\}$ satisfying the third condition of GAC is equivalent to $Pa(Y)\setminus \{W\}$ blocking all generalised back-door paths from $W$ to $Y$.

For any generalised back-door path from $W$ to $Y$ in $\mathcal{M}$, denoted as $\pi$, $\pi$ must be in the form of $W\astleftarrow \dots X \astrightarrow Y$, where $X\in \mathbf{X}$. There are two cases to consider.

\textbf{1}. $\pi$ ends with $X\rightarrow Y$, i.e. $X\in Pa(Y)\setminus \{W\}$. In this case, $X$ is a non-collider on $\pi$, so $\pi$ is blocked by $X$. Moreover, for any other parent of $Y$ in $Pa(Y)\setminus\{W\}$, denoted as $P$, if it is on $\pi$, adding $P$ to the current blocking set (i.e. $\{X\}$) will not open $\pi$, thus $Pa(Y)\setminus\{W\}$ blocks $\pi$.

\textbf{2}. $\pi$ ends with $X\leftrightarrow Y$, i.e. $X\in Sl(Y)\setminus \{W\}$. This means that $X$ is not an ancestor of $Y$ according to Definition~\ref{Ancestral graph} of an ancestral graph and there must be a collider on $\pi$, otherwise there will be a directed path from $X$ to $W$. Then because $\mathcal{M}$ contains $W\rightarrow Y$, $X$ is an ancestor of $Y$, which contradicts the previous statement that $X$ is not an ancestor of $Y$. Furthermore, let $C$ be the closest collider to $X$ on $\pi$. There are two possible cases for $C$. (1) $C=X$, then we have $\astrightarrow C\leftrightarrow Y$. In this case $\forall X^{\prime} \in Pa(Y)\setminus \{W\}$, $X^{\prime}$ must not be in $De(C)$, otherwise $X^{\prime}\in De(C)$ and $X^{\prime}\in Pa(Y)\setminus\{W\}$ such that $C\in An(Y)$, and this contradicts the fact that $C$ (i.e. $X$) is not an ancestor of $Y$. That is, $Pa(Y)\setminus \{W\}$ does not contain a collider on $\pi$ and any descendant of the collider. Thus adding $Pa(Y)\setminus \{W\}$ to the current blocking set (i.e. empty set) still blocks $\pi$. (2) $C\neq X$ and $C$ must be a descendant of $X$, i.e. $\astrightarrow C \leftarrow \dots \leftarrow X\leftrightarrow Y$. In this case, for the same reason as in case (1), $\forall X^{\prime} \in Pa(Y)\setminus \{W\}$, $X^{\prime}$ must not be in $De(C)$. Hence adding $Pa(Y)\setminus \{W\}$ to the current blocking set (i.e. empty set) blocks $\pi$. Therefore, $Pa(Y)\setminus\{W\}$ blocks $\pi$.

Because $Pa(Y)\setminus \{W\}$ blocks any generalised back-door path from $W$ to $Y$ in $\mathcal{M}$, $Pa(Y)\setminus\{W\}$ blocks all generalised back-door paths from $W$ to $Y$, and $Pa(Y)\setminus\{W\}$ is a valid adjustment set relative to $(W, Y)$ in $\mathcal{M}$. Therefore, there exists a set $\mathbf{Z}\subseteq Adj(Y)\setminus\{W\}$ because $Pa(Y)\setminus \{W\} \subseteq Adj(Y)\setminus \{W\}$.

Now we prove that $Pa(Y)\setminus \{W\}$ satisfies the EHS condition with $Q$. Firstly, $Pa(Y)\setminus \{W\}$ satisfies $Q\nindep Y\mid Pa(Y)\setminus \{W\}$ because the path $Q\rightarrow W\rightarrow Y$ is open given any $\mathbf{Z} \subseteq Adj(Y)\setminus \{W\}$. Secondly, $Pa(Y)\setminus \{W\}$ blocks all generalised back-door paths from $W$ to $Y$, so the back-door path starting with $Q\rightarrow W$ must be blocked by $Pa(Y)\setminus \{W\}$. That is, the path $Q\astleftarrow \dots \astrightarrow Y$ is blocked by $Pa(Y)$. The path $Q\rightarrow W\astleftarrow X\dots \astrightarrow Y$ is blocked by $Pa(Y)$ ($W\in Pa(Y)$ is not a blocking node for this path) since $W\astleftarrow X\dots \astrightarrow Y$ is a generalised back-door path from $W$ to $Y$. Moreover, the path $Q\rightarrow W\rightarrow Y$ is blocked by the node $W$. Therefore, all paths between $Q$ and $Y$ are blocked by $Pa(Y)$, i.e. $Pa(Y)$ satisfies $Q\indep Y\mid Pa(Y)$. Hence $Pa(Y)\setminus \{W\}$ satisfies the EHS condition with $Q$. That is, we can conclude that there exists a set $\mathbf{Z}$ in $Adj(Y)\setminus \{W\}$ (\ie  $Pa(Y)\setminus \{W\}$), which satisfies the EHS condition with $Q$ and $\mathbf{Z}$ is a valid adjustment set.
\end{proof}

Theorem~\ref{coro:corCIT} provides the theoretical support for searching for adjustment sets locally using conditional independence tests, directly from data, even when there exist latent variables. Theorem~\ref{coro:corCIT} shows that there exists at least a proper adjustment set in the local causal structure of $Y$; and under the pretreatment assumption, the local causal structure of $Y$ can be discovered from data with latent variables by using a local search algorithm, such as \emph{PC.Select}~\cite{buhlmann2010variable}.  Furthermore, Theorem~\ref{coro:corCIT} guarantees that we can find an adjustment set in the local causal structure around $Y$ when there exists a cause $Q$ of $W$. However, from observational data, it is not always possible to distinguish a cause from a node in $Sl(W)$ (though we may have background knowledge about the causes of $W$). What we can always test is whether there is a variable adjacent to $W$ but not to $Y$. So we relax the requirement for a cause $Q$ of $W$ in Theorem~\ref{coro:corCIT} to be a COSO (Cause of Or Sharing a latent confounder with the treatment Only) variable as defined below.

\begin{definition}[COSO variable]
\label{def:COSO}
A variable $S\in\mathbf{X}$ is a COSO variable \wrt $(W, Y)$  if and only if $S$ is adjacent to $W$ and not adjacent to $Y$ in $\mathcal{M}$. Formally, $S\in Adj(W)\setminus (Adj(Y)\cup \{Y\})$.
\end{definition}

\begin{figure}[t]
\centering
\includegraphics[scale=0.26]{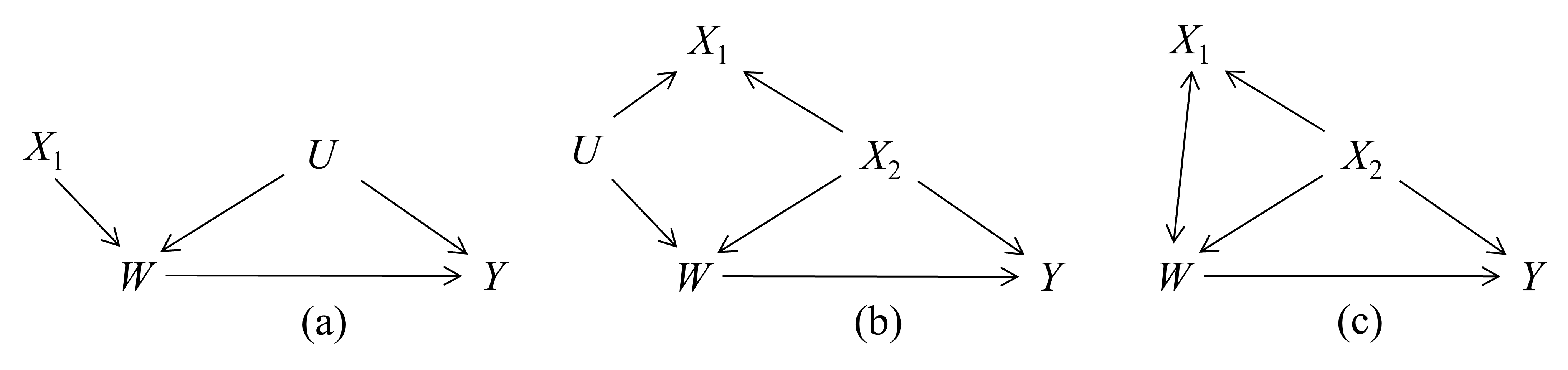}
\caption{An example showing that a COSO variable is not an IV. (a) is a DAG with $U$ being a latent variable and in this DAG $X_1$ is an IV. (b) is a DAG with $U$ being a latent variable and (c) is the corresponding MAG. In the DAG (b), $X_1$ is a COSO variable, but not an IV.}
\label{fig:coso_vs_IV}
\end{figure}

From its definition, we see that a COSO variable can be empirically tested from data, i.e. whether a COSO variable exists is a matter of testing adjacency. It is worth noting that the existence of a COSO variable ensures that $W\rightarrow Y$ is visible. Readers may think that a COSO variable is an instrumental variable (IV)~\cite{bowden1990instrumental,angrist1995two}, which is usually used for estimating causal effect when there exists a latent confounder between $W$ and $Y$. A valid IV, denoted as $X$, must satisfy three conditions: (1) $X$ is a cause of $W$; (2) the causal effect of $X$ on $Y$ is through $W$ only; and (3) there is not a confounding bias between $X$ and $Y$~\cite{hernan2006instruments}. An example IV is shown in Fig.~\ref{fig:coso_vs_IV} (a). The requirements on a COSO variable are more relaxed than those on an IV,  and we show this in the following example.\\

\begin{example}
\label{exp:coso_vs_IV}
In the DAG (a), $X$ is a valid IV \wrt the ordered pair $(W, Y)$. In the MAG in Fig.~\ref{fig:coso_vs_IV} (c), $X_1 \in Adj(W)\setminus(Adj(Y)\cup\{Y\})$, so $X_1$ is a COSO variable based on the Definition~\ref{def:COSO}. However, $X_1$ is not an IV since $X_1$ is not a cause of $W$, and hence the causal effect of $W$ on $Y$ cannot be estimated using $X_1$ as an IV. Instead, the causal effect of $W$ on $Y$ can be estimated by using $\{X_2\}$ as an adjustment set. Both the COSO variable $X_1$ and the adjustment set $\{X_2\}$ can be identified in data. Hence, a COSO variable is not an IV.
\end{example}

An IV and a COSO variable serve for different purposes. An IV is used in the case where the unconfoundedness assumption is not satisfied, whereas a COSO variable is used to estimate causal effects with latent variables when the unconfoundedness assumption is satisfied. Under our problem setting and assuming the existence of a COSO variable, we have the following theorem to identify valid adjustment sets from data and the corollary following it for adjustment set identification with local search.

\begin{theorem} [Adjustment set identification given a COSO variable]
\label{theo:identifiableglobal:coso}
Suppose that there exists a COSO variable $S$, $\mathbf{Z} \subseteq \mathbf{X}\setminus \{S\}$ is a valid adjustment set \wrt $(W, Y)$ if $S\indep Y\mid\mathbf{Z}\cup \{W\}$.
\end{theorem}
\begin{proof}
There are two types of COSO variables, \ie either $S=Q$ or $S\in Sl(W)$. For $S=Q$, Theorem~2 has proved that a valid adjustment set $\mathbf{Z}\subseteq Adj(Y)\setminus\{W\}$ can be identified by $Q\indep Y\mid \mathbf{Z}\cup \{W\}$. Thus, in the following proof, we consider the second type, \ie $S\in Sl(W)$.

The first two conditions of GAC are satisfied by Corollary~\ref{coro:problemsetting}, so if $\mathbf{Z}$ satisfies the third condition of GAC, then $\mathbf{Z}$ is an adjustment set. In other words, we need to prove that if $S\indep Y\mid\mathbf{Z}\cup \{W\}$ holds, then $\mathbf{Z}$ blocks all definite status non-causal paths from $W$ to $Y$, \ie all generalised back-door paths from $W$ to $Y$. We prove this using contradiction. Suppose that a generalised back-door path from $W$ to $Y$, $\pi$, is m-connecting given $\mathbf{Z}$, i.e. $\pi$ is not blocked by $\mathbf{Z}$. There are the following two cases of $\pi$ (depending on whether or not $\pi$ includes $S$): (1) $\pi$ contains $S$ and thus $\pi$ is of the form $W\leftrightarrow S \dots \astrightarrow Y$, then $S$ is m-connected to $Y$ given $\mathbf{Z}\cup \{W\}$, i.e. $S\nindep Y\mid\mathbf{Z}\cup \{W\}$ and this contradicts $S\indep Y\mid\mathbf{Z}\cup \{W\}$. (2) $\pi$ does not contain $S$. Then, the concatenation of the edge $S\leftrightarrow W$ and $\pi$ will produce a path of the form $S\leftrightarrow W\astleftarrow X\dots \astrightarrow Y$, where $X$ is the adjacent node of $W$ on $\pi$. This path is an m-connecting path from $S$ to $Y$ given $\mathbf{Z}\cup \{W\}$ because $S$ is m-connected to $X$ given $W$ (note that $W$ is a collider) and $W$ is m-connected to $Y$ given $\mathbf{Z}$. This means $S\nindep Y\mid\mathbf{Z}\cup \{W\}$, which contradicts $S\indep Y\mid\mathbf{Z}\cup \{W\}$. Thus, if $S\indep Y\mid\mathbf{Z}\cup \{W\}$ holds, $\mathbf{Z}$ blocks all generalised back-door paths from $W$ to $Y$. Therefore, all the three conditions of GAC are satisfied and $\mathbf{Z}$ is a valid adjustment set \wrt  $(W, Y)$.
\end{proof}

Note that Theorem~\ref{theo:identifiableglobal:coso} still requires a global search in $\mathbf{X}$, so for local search, we have the following corollary.
\begin{corollary}
\label{coro:corCIT:coso}
Suppose that there exists a COSO variable $S$, $\mathbf{Z}\subseteq Adj(Y)\setminus\{W\}$ is a valid adjustment set \wrt $(W, Y)$ if $S\indep Y\mid\mathbf{Z}\cup \{W\}$.
\end{corollary}
\begin{proof}
$Pa(Y) \setminus \{W\}$ is a valid adjustment set according to Theorem 2. There must exist at least one valid adjustment set $\mathbf{Z}\subseteq Adj(Y)\setminus\{W\}$ because $Pa(Y)\setminus\{W\} \subseteq Adj(Y)\setminus\{W\}$. From Theorem 3, a set $\mathbf{Z}$ satisfies $S\indep Y\mid\mathbf{Z}\cup \{W\}$, then $\mathbf{Z}$ is a valid adjustment set. Hence, a set $\mathbf{Z}\subseteq Adj(Y)\setminus\{W\}$ satisfies $S\indep Y\mid\mathbf{Z}\cup \{W\}$, then $\mathbf{Z}$ is a valid adjustment set relative to $(W, Y)$.
\end{proof}

Corollary~\ref{coro:corCIT:coso} provides a sound solution for finding an adjustment set locally, \ie once a set $\mathbf{Z}\subseteq Adj(Y)\setminus\{W\}$ satisfies $S\indep Y\mid\mathbf{Z}\cup \{W\}$, then $\mathbf{Z}$ is an adjustment set. We give an example to illustrate Corollary~\ref{coro:corCIT:coso} and its advantages.  
\begin{example}
\label{example01}
Assuming that a dataset is generated from the MAG as in Fig.~\ref{fig:causaldiagram} (b), which contains unobserved variables $X_2$, $X_4$ and $X_6$ of the complete DAG in Fig.~\ref{fig:causaldiagram} (a). In the MAG, $X_1$ or $X_3$ is a COSO variable w.r.t. the visible edge $W\rightarrow Y$. Note that both $X_1$ and $X_3$ can be identified from data using a structure learning algorithm since they are adjacent to $W$ but not to $Y$.

We take $X_1$ as a COSO variable to show how Corollary~\ref{coro:corCIT:coso} works. Corollary~\ref{coro:corCIT:coso} enables us to search for $\mathbf{Z}$ locally in $Adj(Y)\setminus \{W\}  = \{X_5, X_7, X_8, X_9\}$. An adjustment set $\mathbf{Z}$ is found (i.e. $\{X_5, X_9\}$) after searching, in the worst case $2^4$ candidates.  Note that, as we will see late, with our algorithm, the number of candidates searched over is smaller when we use the Apriori strategy to search for a set of minimal adjustment sets. 
\end{example}
Now we differentiate our work from two existing data-driven algorithms, \ie LVIDA and EHS.

\begin{figure}[t]
\centering
\includegraphics[scale=0.31]{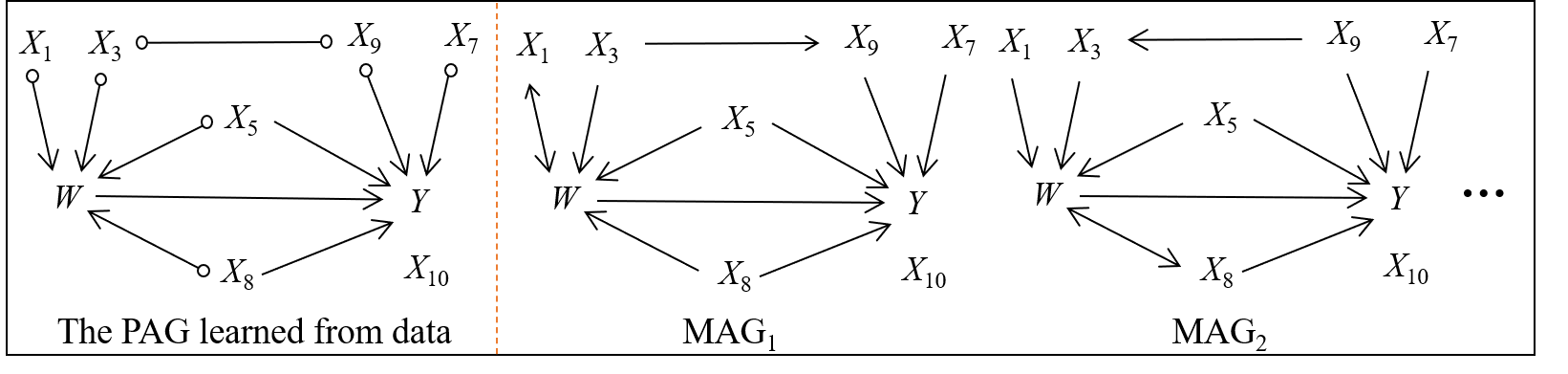}
\caption{The PAG is learned from the data generated from the MAG in Fig 1(b) and two exemplar MAGs inferred from the PAG.}
\label{fig:learnedPAG}
\end{figure}

Our method relies on conditional independence tests, and avoids some biases produced from reading inferred MAGs. LVIDA relies on reading MAGs encoded in a PAG  (partial ancestral graph) learned from data, and may produce biased estimation. A PAG is used to represent a Markov equivalence class of MAGs that share the same set of m-separations~\cite{zhang2008causal,zhang2008completeness}. The PAG learned from the data generated from the MAG in Fig.~\ref{fig:causaldiagram} (b) is shown in Fig.~\ref{fig:learnedPAG}. Two equivalent MAGs inferred from the PAG are also shown in Fig.~\ref{fig:learnedPAG}. When using the two MAGs, node $X_8$ will be included in the adjustment set based on the generalised back-door path criterion~\cite{maathuis2015generalized} or the GAC (Definition~\ref{def:GAC})~\cite{perkovic2018complete}. This has been shown by the poor performance of LVIDA with the synthetic datasets in our experiments (see Section~\ref{app:subsec:syndata}). In contrast, our method will not include $X_8$ since conditional independence tests are used, instead of inferring MAGs from the PAG. Note that in the DAG in Fig.~\ref{fig:causaldiagram} (a), an $M$-structure~\cite{scutari2009learning,greenland2003quantifying,pearl2009myth} forms at $X_8$, which causes the well-known Berkson's paradox~\cite{berkson1946limitations}. In general, an $M$-structure contains the path $X\leftarrow U_1 \rightarrow M\leftarrow  U_2 \rightarrow X'$ where $X$, $M$ and $X'$ are observed, but $U_1$ and $U_2$ are unobserved~\cite{scutari2009learning,greenland2003quantifying,pearl2009myth}. Conditioning on $M$ generates a spurious association between $X$ and $X'$. Therefore, in the example in Fig.~\ref{fig:causaldiagram} (a) adjusting on (i.e. conditioning on) $X_8$ when $X_2$ and $X_6$ are unobserved will produce a biased estimation for $ACE(W, Y)$. Our method does not suffer from $M$-bias whereas other data-driven methods may because our method does conditional independence tests to determine proper adjustment sets from data directly instead of reading an adjustment set from a learned PAG.

The local search in Corollary~\ref{coro:corCIT:coso} reduces the search complexity of searching for the adjustment set with respect to a COSO variable from $O(2^{p-1})$ to $O(2^{q})$ where $q$ is the size of $Adj(Y)\setminus \{W\}$. Note that $p$ is the number of observed variables in a DAG and is much larger than $q$. EHS also uses conditional independence tests to search for adjustment sets but does not restrict candidates to be local. The local search by using Corollary~\ref{coro:corCIT:coso} is in theory incomplete to the EHS condition, in the sense that there may exist an adjustment set that satisfies the EHS condition but cannot be found by local search. We provide an example to show this.

\begin{figure}[t]
\centering
\includegraphics[scale=0.28]{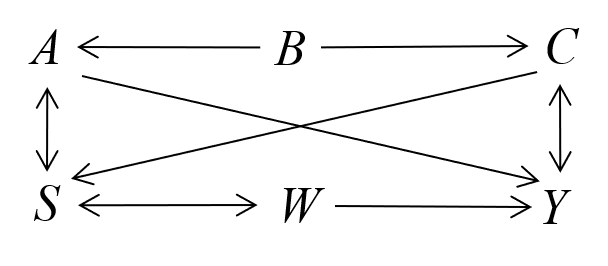}
\caption{An example showing that Corollary~\ref{coro:corCIT:coso} fails to pick up an adjustment set by local search.}
\label{fig:counterexample}
\end{figure}

\begin{example}
\label{lab:exp:002}
Refer to the MAG in Fig.~\ref{fig:counterexample}, denoted as $\mathcal{M}$, which contains a pair of variables $(W, Y)$, and $\mathbf{X} = \{S, A, B, C\}$, which are all pretreatment variables. Note that $S$ is a COSO variable and $W\rightarrow Y$ is visible in $\mathcal{M}$.

For every $\mathbf{Z}\subseteq Adj(Y)\setminus \{W\}$, we have $S\nindep Y \mid \mathbf{Z}\cup \{W\}$ since adjusting for $\{A, C\}$ only leaves
the path $S\leftrightarrow A\leftarrow B\rightarrow C\leftrightarrow Y$ open in $\mathcal{M}$. We cannot find an adjustment set locally in $Adj(Y)\setminus \{W\}$. However, there is a global adjustment set, which is $\{A, B, C\}$ in this MAG based on the EHS condition.   
\end{example}

\noindent \textbf{Remark}: A generalisation of the above counterexample is that the back-door path $\pi$ from $W$ to $Y$ consists of $W\leftrightarrow S \astleftarrow \dots \astrightarrow C\leftrightarrow Y$ and each collider on $\pi$ is an ancestor of either $S$ or $Y$. In this case, a non-collider node on $\pi$ can be used to block $\pi$, but locally, we cannot find such a non-collider node based on Corollary~\ref{coro:corCIT:coso}. The incompleteness of the local search of Corollary~\ref{coro:corCIT:coso} is the result of using a variable in $Sl(W)$ as a COSO variable. When a user knows there is a cause of $W$, the local search is complete. However, without such knowledge, a cause and a variable in $Sl(W)$ cannot be distinguished from data, and the incompleteness is a limitation of a data-driven algorithm.

We have conducted a simulation in Section~\ref{subsec:completeness} to show that these cases are rare. In the simulation, we check the completeness of Corollary~\ref{coro:corCIT:coso} and the result shows that the number of MAGs where Corollary~\ref{coro:corCIT:coso} cannot find an adjustment set locally is small. Although Corollary~\ref{coro:corCIT:coso} is incomplete, its coverage is close to 100\% in most situations.

To reduce the random fluctuations without sacrificing much efficiency, in our proposed $CEELS$ algorithm, we extend the search space of Corollary~\ref{coro:corCIT:coso} to $Adj(W\cup Y)\setminus\{W, Y\}$, where $Adj(W\cup Y)$ is a short form of $Adj(W)\cup Adj(Y)$. Our experiments show that the extended search space makes it more likely to find an adjustment set by local search. However, in $Adj(W\cup Y)\setminus\{W, Y\}$, there might be multiple adjustment sets and randomly using one of the adjustment sets can cause random fluctuations in causal effect estimation~\cite{textor2011adjustment}. Therefore we use each of the adjustment sets found to obtain an estimation of the causal effect and then average over all the estimated causal effects as the output of the proposed $CEELS$ algorithm.

Furthermore, a large sized adjustment set requires more data samples for reliable statistical tests~\cite{de2011covariate}, so we aim to find all minimal adjustment sets in $Adj(W\cup Y)\setminus\{W, Y\}$.
\begin{definition}[Minimal adjustment set]
\label{minimalset}
$\mathbf{Z}$ is a minimal adjustment set if no proper subset of $\mathbf{Z}$ is an adjustment set.
\end{definition}

We continue to use Example~\ref{example01} to illustrate the advantage of minimal adjustment sets. In the MAG of Fig.~\ref{fig:causaldiagram} (b), the valid adjustment sets include $\{X_3, X_5\}$, $\{X_5, X_9\}$, $\{X_1, X_3, X_5\}$, $\{X_1, X_5, X_9\}$, $\dots$, $\{X_1, X_3, X_5, X_7, X_9, X_{10}\}$. The number of valid adjustment sets is $24$. Finding all 24 adjustment sets is computationally expensive, but it can be even worse as using all the adjustment sets for causal effect estimation may result in unreliable causal effect estimation since an adjustment set with many variables will need a large dataset to ensure statistic reliability in the causal effect estimation. However, with our method, we can find the minimal adjustment sets, $\{X_3, X_5\}$ and $\{X_5, X_9\}$. The search is efficient and the requirement of sample size is much lower.

\subsection{The $CEELS$ Algorithm}
\label{subsec:pracalg}
In this section, we present the details of the $CEELS$ (Causal Effect Estimation by Local Search) algorithm. To achieve high efficiency, in addition to local search, $CEELS$ employs a frequent pattern mining approach (the \emph{Apriori} approach)~\cite{agrawal1994fast} to conduct a level-wise search in $Adj(W\cup Y)\setminus\{W, Y\}$ to discover minimal adjustment sets. The level-wise search is based on the fact that a superset of a minimal adjustment set is not a minimal adjustment set. Hence, the process of discovering minimal adjustment sets has the upward closure property and thus we can use the Apriori~\cite{agrawal1994fast} approach to prune the search space based on the upward closure property of minimal adjustment sets to improve the efficiency of the search for minimal adjustment sets.

\begin{algorithm}[t]
\caption{Causal Effect Estimation by Local Search ($CEELS$)}
\label{pseudocode01}
\noindent {\textbf{Input}}: Dataset $\mathbf{D}$ with the treatment $W$, pretreatment variables $\mathbf{X}$, the outcome $Y$.\\
\noindent {\textbf{Output}}: $ACE(W, Y)$.
\begin{algorithmic}[1]
	\STATE {call a local structure learning algorithm to search for $Adj(W)$ and $Adj(Y)$ from $\mathbf{D}$.}
	\STATE {$Adj'(W\cup Y)= Adj(W)\cup Adj(Y)\setminus\{W, Y\}$}
	\STATE{$\mathbf{COSO} =  Adj(W)\setminus (Adj(Y)\cup \{Y\})$}
	\STATE{let $\mathbf{\psi} = \emptyset$; \\ /*Search for minimal adjustment sets, and calculate $ACE(W, Y)$*/}
	\FOR{each variable $S\in \mathbf{COSO}$}
	\STATE {$\mathbf{C}_1 = Adj'(W\cup Y)\setminus \{S\}$}
	\STATE {$k = 1$}
	\WHILE{$\mathbf{C}_{k}\neq \emptyset$}
	\FOR {each element $\mathbf{Z}\in \mathbf{C}_{k}$}
	\IF{$\mathbf{Z}$ is not in $\mathbf{\psi}$}
	\IF {$S\indep Y\mid\mathbf{Z}\cup \{W\}$}
	\STATE{calculate $ACE(W, Y)$ via Eq.(\ref{eq002}) and remove $\mathbf{Z}$ from $\mathbf{C}_{k}$}
	\STATE{$\mathbf{\psi}\leftarrow \cup(\mathbf{Z}, ACE(W, Y))$}
	\ENDIF
	\ENDIF
	\ENDFOR
	\STATE{$k = k+1$}
	\STATE{$\mathbf{C}_{k} = Candidate.gen(\mathbf{C}_{k-1}, k)$ /*Apriori pruning strategy*/}
	\ENDWHILE
	\ENDFOR
	\RETURN{$\mathbf{\psi}$}
\end{algorithmic}
\noindent {\textbf{Function}: $Candidate.gen$($\mathbf{C}_{k-1}$; $k$)}\\
\noindent {/*Generate candidate adjustment sets $\mathbf{C}_k$ based on the sets in $\mathbf{C}_{k-1}$*/}
\begin{algorithmic}[1]
	\STATE {$\mathbf{C}_k = \emptyset$;}
	\STATE {Hash the first $(k-2)$ variables of all sets in $\mathbf{C}_{k-1}$}
	\FOR {each pair $C_i, C_j\subseteq {\mathbf{C}_{k-1}}$ with the same first $(k-2)$ variables.}
	\STATE{$C_s = \cup({C_i}, {C_j})$ }
	\STATE {$Csubsets=subset(C_s,k-1)$ /*Generating all subsets of $C_s$ with the size $k-1$.*/}
	\IF {every element of $Csubsets$ is in $\mathbf{C}_{k-1}$}
	\STATE {Add $C_s$ to $\mathbf{C}_k$}
	\ENDIF
	\ENDFOR
	\RETURN{$\mathbf{C}_k$}
\end{algorithmic}
\end{algorithm}

$CEELS$ (presented in Algorithm~\ref{pseudocode01}) aims to identify the minimal adjustment sets in $Adj'(W\cup Y)=Adj(W\cup Y)\setminus\{W, Y\}$ and then estimate the causal effect of $W$ on $Y$ ($ACE(W, Y)$) by adjusting for a minimal adjustment set found. Each of identified minimal adjustment set $\mathbf{Z}$ and the estimated causal effect $ACE(W, Y)$ is stored in $\mathbf{\psi}$. We utilise the level-wise search strategy to prune the search space. A set of candidate adjustment sets are generated using the function  $Candidate.gen(\mathbf{C}_{k-1}; k)$.

As shown in Algorithm~\ref{pseudocode01}, $CEELS$ firstly (Lines 1 to 3) searches for $Adj'(W\cup Y)$ and the set of COSO variables, $\mathbf{COSO} = Adj(W)\setminus (Adj(Y)\cup \{Y\})$. Line 1 finds $Adj(W)$ and $Adj(Y)$, respectively, based on a local causal structure learning algorithm, such as \emph{PC.Select}~\cite{buhlmann2010variable}, \emph{MMPC} and \emph{HITON-PC}~\cite{aliferis2010local}. In our implementation, \emph{PC.Select} algorithm is employed. Then, for each variable $S$ in the $\mathbf{COSO}$ set, Lines 5 to 20 aim to search for the minimal adjustment sets in $Adj'(W\cup Y)\setminus \{S\}$. Line 9 enumerates every $\mathbf{Z}$ in the set of $\mathbf{C}_k$ which contains candidates of size $k$. Line 10 is to prevent repeated tests for the adjustment sets that are already in $\mathbf{\psi}$ using a different COSO variable. The conditional independence test in Line 11 is to assess whether a candidate set $\mathbf{Z}$ is an adjustment set or not according to Corollary 2. In Line 12, $ACE(W, Y)$ is estimated based on Eq.(2) by adjusting for the current minimal adjustment set $\mathbf{Z}$.

The function $Candidate.gen(\mathbf{C}_{k-1}; k)$ is used in Line 18 to generate level $k$ candidate adjustment sets. $Candidate.gen(\mathbf{C}_{k-1}; k)$ uses the upward closure property (Apriori approach~\cite{agrawal1994fast}) to prune the search space when generating candidate minimal adjustment sets in a level-wise search. Once a minimal adjustment set $\mathbf{Z}$ is found at level $k$, none of its supersets will be generated as a level $l$ $(l > k)$ candidate adjustment set. We employ a \emph{hash table} to store candidates to save search time (See Line 2 in $Candidate.gen(\mathbf{C}_{k-1}; k)$). All variables are sorted in lexicon order for fast pruning. The first $(k-2)$ variables of all sets in $\mathbf{C}_{k-1}$ are hashed for generating new candidates in $\mathbf{C}_{k}$ (See Line 3). Line 4 of the function combines a pair of sets with the same first $(k-2)$ variables to create a new candidate with $k$ variables. Lines 5 to 9 prune candidate set $\mathbf{C}_{k}$ by checking all subsets of a candidate to ensure the candidate is potentially minimal. 

\noindent \textbf{Time complexity.} The time complexity of $CEELS$ comes from two sources, the complexity of \emph{PC.select}~\cite{buhlmann2010variable} and the complexity of the \emph{Apriori} algorithm~\cite{agrawal1994fast}. A crude complexity bound of \emph{PC.select} is $\mathbf{O}(p*2^{q}*n))$ where $p$ and $n$ are the numbers of variables and samples respectively and $q=\max(|Adj(W)\setminus\{Y\}|, |Adj(Y)\setminus\{W\}|)$. The complexity for the first part is $\mathbf{O}(p*2^{q}*n)$. For the Lines 5 to 20 of $CEELS$, the number of iterations of the for loop is at most $p$, and the number of iterations of the while loop from Lines 8 to 19 is determined the time complexity of the \emph{Apriori} algorithm~\cite{agrawal1994fast}. Theoretically, its time complexity is $\mathbf{O}(2^{l})$ where $l$ denotes the sizes of $Adj^{'}(W\cup Y)\setminus \{S\}$. Because of the pruning, $l$ is a small number. We let $l=q$ for simplicity. In the worst scenario, the time complexity of the second part is $\mathbf{O}({p}*2^{q})$. Hence, the overall complexity of $CEELS$ is $\mathbf{O}(p*2^{q}*n)$, which is largely determined by the number of the neighbour variables around $W$ and $Y$. In practice, the largest size of the conditional set for conditional independence tests in \emph{PC.select} can be set to a small number, $r$~\cite{buhlmann2010variable,aliferis2010local}. The size of the largest minimal adjustment set is also set to a small number $r$. The complexity is $\mathbf{O}(p*2^{r}*n)$ where $r$ is a small number, typically 2 to 5.

\section{Experiments}
\label{sec:exp}
\subsection{Comparison Methods and Implementations}
\label{subsec:comparsion}
Nine benchmark methods/criteria are used in the comparison, including the pretreatment adjustment (PRE for short)~\cite{rubin1974estimating}, the four minimal adjustment set selection (MASS) criteria~\cite{de2011covariate} $\hat{\mathbf{X}}_{\rightarrow W}$, $\hat{\mathbf{X}}_{\rightarrow Y}$, $\hat{\mathbf{Q}}_{\rightarrow W}$ and $\hat{\mathbf{Z}}_{\rightarrow Y}$, the disjunctive cause criterion~\cite{vanderweele2011new} $\hat{\mathbf{X}}_{\rightarrow W, Y}$, EHS~\cite{entner2013data}, and LVIDA with FCI (LVIDA-FCI), LVIDA with rFCI (LVIDA-rFCI)~\cite{malinsky2017estimating}. EHS returns a set of unbiased estimates and uses the average as its final output. LVIDA provides a set of possible estimates and also uses the average as its final result. Note that if the edge between $(W, Y)$ is learned as $W\leftrightarrow Y$ by FCI (or rFCI), LVIDA-FCI (or LVIDA-rFCI) returns $ACE(W, Y)=0$.  Details of these methods/criteria and their implementations are provided below. 

\subsubsection{Details of Comparison Methods}
The details of the comparisons are introduced as follows:
\begin{enumerate}
\item PRE~\cite{rubin1974estimating}, adjusts for all pretreatment variables $\mathbf{X}$ for estimating $ACE(W, Y)$.
\item The MASS criteria refer to the four criteria used to select the minimal adjustment set from pretreatment variables as summarised in~\cite{de2011covariate}. The four criteria select all causes of $W$, all causes of $Y$, all causes of $W$ excluding causes of $Y$, and all causes of $Y$ excluding causes of $W$, as the adjustment set respectively. Following the notation in~\cite{de2011covariate}, the identified adjustment sets are denoted as $\hat{\mathbf{X}}_{\rightarrow W}$, $\hat{\mathbf{X}}_{\rightarrow Y}$, $\hat{\mathbf{Q}}_{\rightarrow W}$ and $\hat{\mathbf{Z}}_{\rightarrow Y}$ respectively.
\item The disjunctive cause criterion~\cite{vanderweele2011new} considers the union of all causes of $W$ and all causes of $Y$ as the adjustment set, denoted as $\hat{\mathbf{X}}_{\rightarrow W,Y} = (\hat{\mathbf{X}}_{\rightarrow W} \cup \hat{\mathbf{X}}_{\rightarrow Y})$.
\item EHS~\cite{entner2013data}, a data-driven method for adjustment set identification from data with latent variables (based on Theorem~\ref{Theo:EHSsearch}), using conditional independence test and the pretreatment assumption.
\item LVIDA~\cite{malinsky2017estimating} is a causal effect estimation algorithm based on global causal structure learning. The LVIDA method using FCI for causal structure learning is denoted as LVIDA-FCI, and the LVIDA method using rFCI (the really Fast Causal Inference, rFCI~\cite{colombo2012learning}) is called LVIDA-rFCI. 
\end{enumerate}

\subsubsection{Details of Implementation \& Parameter Setting}
\textbf{Calculating $ACE(W, Y)$ by covariate adjustment}. For all methods, linear regression and logistic regression are used to estimate the average causal effect of $W$ on $Y$ with covariate adjustment, i.e. $ACE(W, Y)$ as given in Eq.(2) for continuous and binary outcomes, respectively. With linear regression, the causal effect is calculated as the partial regression coefficient of $W$. It has been shown that when the adjustment set is right, the coefficient of $W$ is the $ACE(W, Y)$~\cite{malinsky2017estimating}. For binary outcome, the marginal causal odds ratio (MCOR) as in Eq.(\ref{eq005}) is the estimated $ACE(W, Y)$. Logistic regression is employed to obtain $\log(MCOR)$, and this is implemented by an $\mathbf{R}$ package \emph{stdReg}~\cite{sjolander2016regression}.
\begin{equation}
\label{eq005}
MCOR  = \frac{P(y=1,do(w=1))/P(y=0,do(w=1))}{P(y=1,do(w=0))/P(y=0,do(w=0))}
\end{equation}

For implementing $CEELS$, \emph{PC.select} from the $\mathbf{R}$ package \emph{pcalg} is employed to learn the local causal structure around $W$ and $Y$, i.e. $Adj(W)$ and $Adj(Y)$. The conditional independence tests are implemented by using $gaussCItest$ and $binCItest$ from the same $\mathbf{R}$ package \emph{pcalg}. The average value of these $ACE(W, Y)$ estimated by $CEELS$ is regarded as the final result.

PRE considers the set $\mathbf{X}$ (all pretreatment variables) as the adjustment set for estimating $ACE(W, Y)$, so we use the $CEELS$ algorithm without adjustment variable selection to implement it. The implementations for MASS and the disjunctive cause criterion are from the $\mathbf{R}$ package \emph{CovSelHigh}~\cite{haggstrom2018data}. The parameter setting are \emph{method=mmpc}, \emph{simulate=FALSE}, \emph{betahat=FALSE}.

EHS was originally implemented in Matlab~\footnote{\url{https://sites.google.com/site/dorisentner/publications/CovariateSelection}}~\cite{entner2013data}.  We implement the $\mathbf{R}$ version of EHS as the Matlab implementation is for the linear Gaussian model only. As with $CEELS$, we utilise $gaussCItest$ and $binCItest$ to implement the conditional independence tests for EHS. The maximum size of a conditioning set is set to 6 for EHS in all experiments as it is impractical to use large conditioning sets (although as per its design, EHS conducts conditional independence tests over all possible conditioning sets).  As the output of EHS contains multiple estimated values for $ACE(W, Y)$, each corresponding to a valid adjustment set identified by EHS, we use the average of the multiple causal effect values estimated by EHS as its final result in the comparison.

For LVIDA, we use the implementation by the authors~\cite{malinsky2017estimating}\footnote{\url{https://github.com/dmalinsk/lv-ida}}, and the functions FCI and rFCI are from the $\mathbf{R}$ package \emph{pcalg}. The function \emph{lv.ida} choose \emph{method=``local''} for ensuring the fast search of valid adjustment set $\mathbf{Z}$. For the binary outcome, we use \emph{stdReg} to replace the linear regression. As LVIDA is a bound estimation method that outputs a multiset of causal effect values, each corresponding to a potential (not necessarily proper) adjustment set, we use the average of these estimated values as the result of LVIDA in the comparison. The main drawback of LVIDA is that it produces a bound estimation of causal effects since a PAG encodes a Markov equivalence class of MAGs as mentioned in the Introduction section.

\noindent \textbf{Evaluation metric}. Relative Error (RE, \%) is the absolute error of the estimated causal effect relative to the true causal effect (in percentage). RE is used to evaluate the estimated bias of the estimators. In all experiments, the significance level ($\alpha$) is set to 0.05 for all algorithms involved. 

\subsection{Evaluation on Synthetic Data}
\label{app:subsec:syndata}
\begin{figure}[t]
\centering
\includegraphics[scale=0.28]{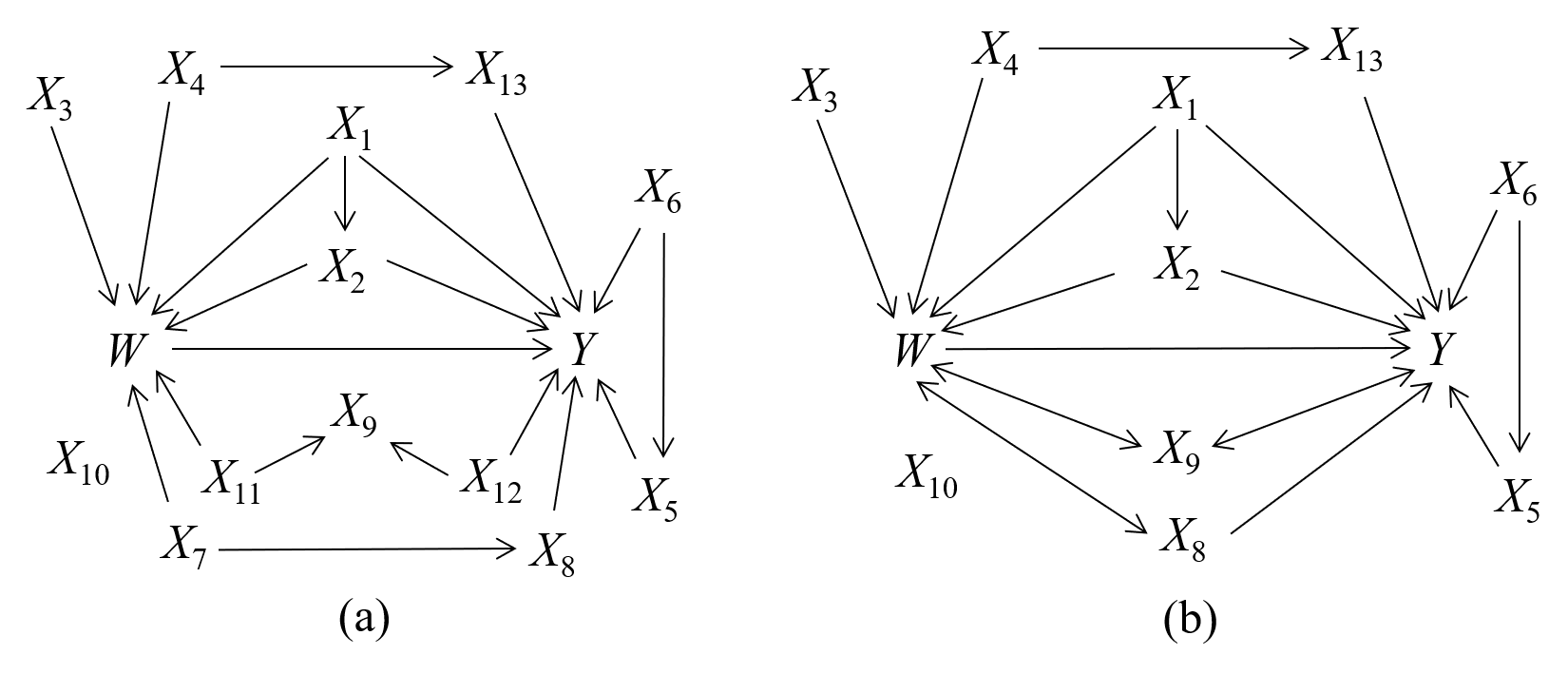}
\caption{(a) A causal DAG for synthetic dataset generation~\cite{haggstrom2018data} and its corresponding MAG in Fig. (b) where $X_7, X_{11}, X_{12}$ are removed.}
\label{fig:Syn_fig02}
\end{figure}

The DAG in Fig.~\ref{fig:causaldiagram}(a)~\cite{witte2019covariate} contains 12 variables and the DAG in Fig.~\ref{fig:Syn_fig02}(a)~\cite{haggstrom2018data} includes 15 variables. The DAG in Fig.~\ref{fig:causaldiagram}(a) has three back-door paths between $W$ and $Y$ (a sparse structure around $W$ and $Y$) and the DAG in Fig.~\ref{fig:Syn_fig02}(a) has seven back-door paths between $W$ and $Y$ (a denser structure around $W$ and $Y$ than the DAG in Fig.~\ref{fig:causaldiagram}(a)). The corresponding MAGs are provided in Fig.~\ref{fig:causaldiagram}(b) and~\ref{fig:Syn_fig02}(b), respectively. Both true MAGs contain an $M$-structure, which causes the $M$-bias~\cite{greenland2003quantifying,pearl2009myth}. In order to assess the ability of $CEELS$ to deal with latent variables, all synthetic datasets are generated with latent variables and $M$-bias.

We firstly generate two groups of synthetic datasets without latent variables based on the two true DAGs in Fig.~\ref{fig:causaldiagram}(a) and Fig.~\ref{fig:Syn_fig02}(a), respectively. We follow the processes for synthetic data generation in the literature, \ie \cite{witte2019covariate} and~\cite{haggstrom2018data} respectively. Secondly, we generate synthetic datasets with latent variables by removing several common causes from the above generated datasets based on Fig.~\ref{fig:causaldiagram}(a) and Fig.~\ref{fig:Syn_fig02}(a), and denote them as Group I datasets and Group II datasets. Each group includes 14 datasets with a range of sample sizes, i.e. 1k (stands for 1,000), 2k, 3k, 4k, 5k, 6k, 7k, 8k, 9k, 10k, 30k, 50k, 100k, and 150k. From each of the datasets, the columns for variables $X_2$, $X_4$ and $X_6$ of Group I datasets based on the DAG in Fig.~\ref{fig:causaldiagram}(a) and the columns for $X_7, X_{11}$ and $X_{12}$ of Group II datasets based on the DAG in Fig.~\ref{fig:Syn_fig02}(a) are removed. The resulted datasets correspond to MAGs in Fig.~\ref{fig:causaldiagram}(b) and Fig.~\ref{fig:Syn_fig02}(b) respectively.

The true causal effect of all synthetic datasets in Group I is 0.5~\cite{witte2019covariate} and the true causal effect of all synthetic datasets in Group II is 2~\cite{haggstrom2018data}. In order to avoid the bias caused by data generation, we repeat the procedure 50 times and generate 50 datasets in each setting. Finally, we conduct experiments on these datasets with latent variables to verify the performance of $CEELS$ for the task of estimating causal effects. We repeat the experiments 50 times and the REs (average of 50 results) of each of the methods/criteria on all synthetic datasets are reported in Fig.~\ref{Result:Nocausalsuff:samples}. For the standard deviations, please see Tables 1 and 2 in Supplement.

\begin{figure}[t]
\begin{minipage}{1\linewidth}\centering
	\includegraphics[scale=0.276]{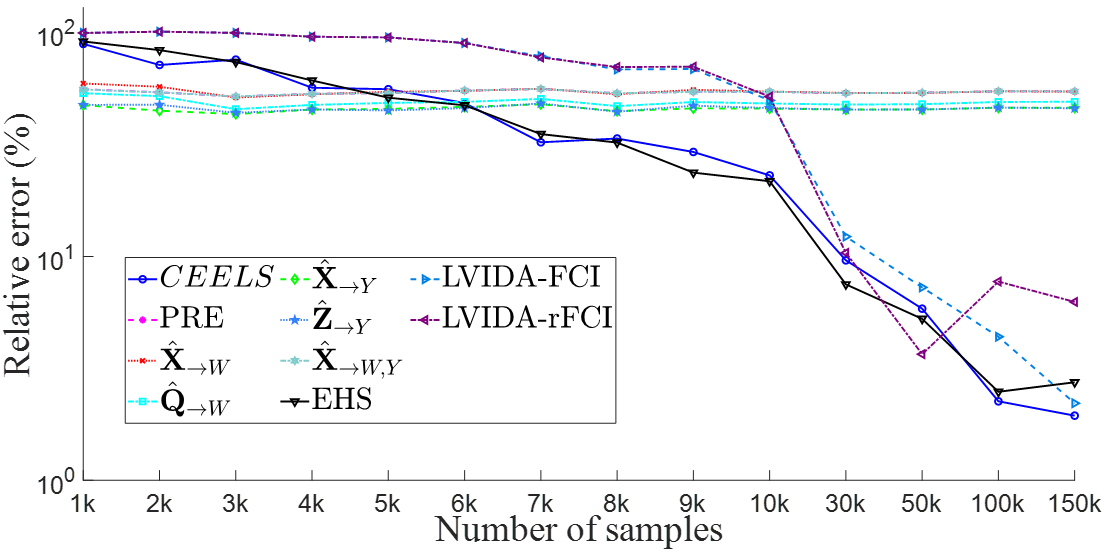}
\end{minipage}%
\vfil
\begin{minipage}{1\linewidth}\centering
	\includegraphics[scale=0.276]{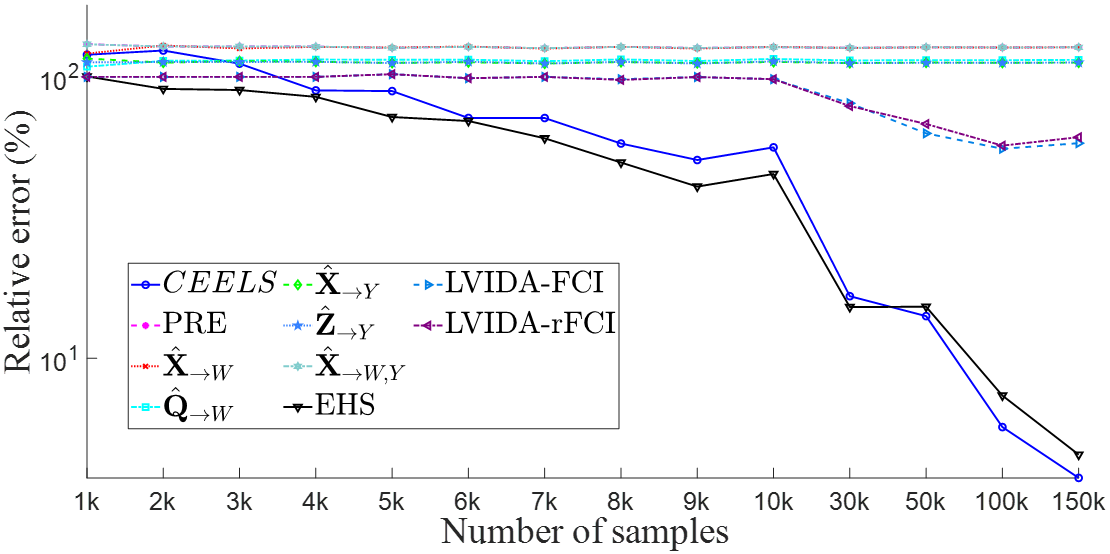}
\end{minipage}%
\caption{REs on synthetic datasets with increasing sample size ($\log_{10}$ on the $Y$-axis). The top panel shows the results of Group I datasets, and the bottom panel shows the results of Group II datasets. As the sample size grows, the REs of $CEELS$ and EHS decrease significantly. The performances of LVIDA-FCI and LVIDA-rFCI are worse than $CEELS$.}
\label{Result:Nocausalsuff:samples}
\end{figure}

From Fig.~\ref{Result:Nocausalsuff:samples}, we have the following observations: (1) The methods which assume causal sufficiency, including PRE, $\hat{\mathbf{X}}_{\rightarrow W}$, $\hat{\mathbf{Q}}_{\rightarrow W}$, $\hat{\mathbf{X}}_{\rightarrow Y}$, $\hat{\mathbf{Z}}_{\rightarrow Y}$ and $\hat{\mathbf{X}}_{\rightarrow W, Y}$ have a large relative error on both groups of datasets because they do not deal with latent variables; (2) The performances of $CEELS$ and EHS are consistent and improve with the increase of sample size; (3) In most cases, the performances of LVIDA-FCI and LVIDA-rFCI are worse than $CEELS$ and EHS since LVIDA-FCI and LVIDA-rFCI rely on the correctness of the learned global causal structure.

The poor performances of LVIDA-FCI and LVIDA-rFCI on Group II datasets reinforce our argument in the Introduction that there is a need for causal effect estimation using local causal structure learning. The complexity for optimally learning a global structure increases exponentially with the number of variables, and the current PAG learning methods are error-prone in edge orientations. LVIDA-FCI and LVIDA-rFCI each obtain a possible set of adjustment sets from a learned PAG, and both depend heavily on the quality of the PAG.  This has been demonstrated by the (better) performances of EHS and $CEELS$, both having not used heuristics, but EHS is inefficient as a result of global search, and  $CEELS$ is efficient due to the local search.  

\subsection{Evaluation Using a Benchmark Bayesian Network}
\label{subsec:bayesiannetwork}
To verify the correctness of $CEELS$, we generated synthetic datasets with latent variables based on a benchmark Bayesian network, INSURANCE, which contains 27 nodes and 52 arcs\footnote{The network can be downloaded at \url{http://www.bnlearn.com/bnrepository/discrete-medium.html}}. We select the node \emph{Cushioning} as the treatment variable $W$ and the node \emph{MedCost} as the outcome variable $Y$ such that all other variables are pretreatment variables, i.e. there are no descendant nodes of $W$ and $Y$ in the INSURANCE network.

In this experiment, we will use the underlying DAG to calculate the ground truth causal effect since there are no ground truths given. Therefore, it is essential that the generated datasets are faithful to the DAG used for generating datasets without latent variables. We use a large sample size when generating the synthetic datasets and we ensure the DAG can be learned back from a generated dataset as follows. For each generated dataset, the \emph{PC} algorithm~\cite{spirtes2000causation} (a global structure learning algorithm) is used to learn a CPDAG (completed partially directed acyclic graph) from the data. If the learned CPDAG is Markov equivalent to the DAG based on which the dataset is generated, then the generated dataset is kept for the experiment. 
Otherwise, the dataset is discarded.

We generate 7 synthetic datasets with 200k, 250k, 300k, 350k, 400k, 450k and 500k samples, respectively, based on the INSURANCE network. The ground truth causal effect $ACE(W, Y)$ for each dataset is calculated by adjusting for variables \emph{Airbag} and \emph{RuggedAuto}, which are identified by using the back-door criterion~\cite{pearl1995causal}. We then produce 7 datasets with latent variables by removing four variables from the above generated datasets, which are \emph{VehicleYear}, \emph{MakeModel}, \emph{RuggedAuto} and \emph{Accident}. The resulting MAG contains an $M$-structure $W \leftrightarrow ThisCarDam \leftrightarrow Y$, which means all datasets produced contain $M$-bias. All experiments are conducted on these datasets with latent variables.

\begin{figure}[t]
\centering
\includegraphics[scale=0.28]{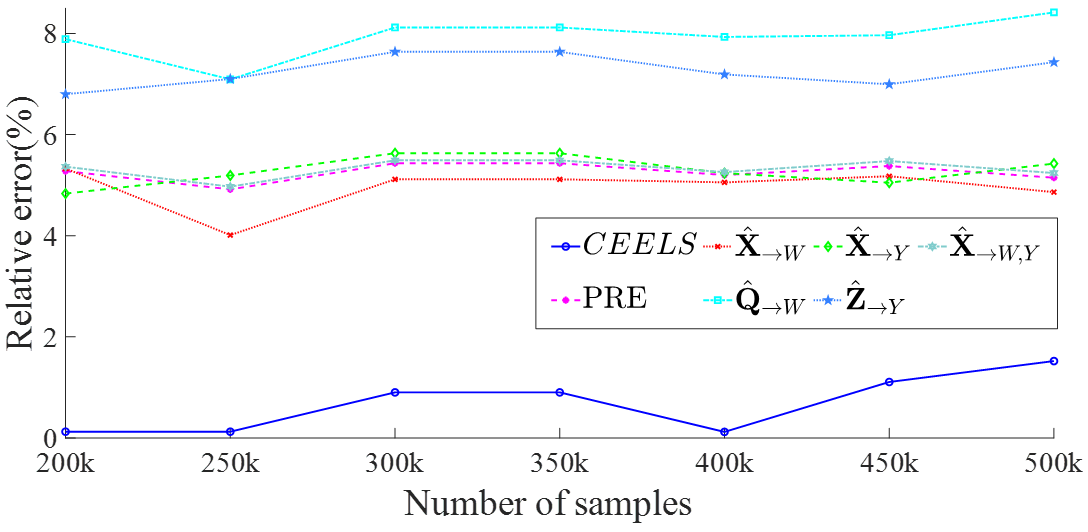}
\caption{REs of causal effects estimated by different methods with the datasets generated based on INSURANCE. EHS, LVIDA-FCI, and LVIDA-rFCI are not shown since they did not return results within two hours. $CEELS$ outperforms the six other adjustment criteria/methods.}
\label{ResultBN:insufficient}
\end{figure}

For the 7 synthetic datasets with latent variables, EHS, LVIDA-FCI and LVIDA-rFCI did not return results within two hours due to their low efficiency. Hence, we only compare $CEELS$ with the other six methods and the results are shown in Fig.~\ref{ResultBN:insufficient}. $CEELS$ outperforms all the six methods for causal effect estimation and this is expected since the six methods require causal sufficiency.

\subsection{Experiments with Real-world Datasets}
\label{subsec:realworld}
We further evaluate the performance of $CEELS$ using two real-world datasets: IHDP~\cite{hill2011bayesian} and Twins~\cite{louizos2017causal}, which are widely used benchmarks for causal inference. 

\textbf{The Infant Health and Development Program (IHDP) dataset}~\cite{hill2011bayesian} is from a collection based on a randomised controlled experiment that studied high-quality intensive care provided to low-birth-weight and premature infants. There are  747 infants, including 139 treated infants and 608 control infants, each with 24 pretreatment variables (excluding race). Children who did not receive specialist visits formed a control group. The outcome was simulated from the $\mathbf{R}$ package \emph{npci}\footnote{\url{https://github.com/vdorie/npci}} such that we have the true $ACE(W, Y)= 4.36$ as in the work~\cite{hill2011bayesian}.

\textbf{The Twins dataset} is a benchmark dataset that comes from the recorded twin births and deaths in the USA between 1989 -1991~\cite{almond2005costs}. We only choose same-sex twins with weights less than 2000g from the original data, and each twin-pair contains 40 pretreatment variables related to the parents, the pregnancy, and the birth~\cite{louizos2017causal}. For each record, both the treatment $W$=1 (heavier twin) and $W$=0 (lighter twin) are observed. The mortality after one year is the true outcome, and the ground truth $ACE(W, Y)$ is -0.02489. We eliminate all records with missing values and have 4821 twin-pairs left. For simulating an observational study, following the work in~\cite{louizos2017causal}, we use Bernoulli distribution to randomly hide one of the two twins, that is $W_i\mid\mathbf{x}_i\sim Bern(sigmoid(\beta^{T}\mathbf{x}+\varepsilon))$, where $\beta^{T}\sim\mathcal{U}((-0.1,0.1)^{40\times1})$ and $\varepsilon\sim\mathcal{N}(0,0.1)$.

The comparison results of all algorithms are listed in Table~\ref{tab:IHDP_twins}. For both datasets, $CEELS$ has achieved the lowest REs of 3.49\% and 5.66\%, respectively. As the datasets were collected from well designed observational studies, and the covariates were chosen by domain experts, they satisfy causal sufficiency and pretreatment assumptions. Hence all methods achieved good performance.

However, $CEELS$ stands out because it finds the minimal adjustment sets. EHS has a larger RE compared to $CEELS$ because most of the returned adjustment sets contain multiple redundant variables. Take the IHDP dataset as an example the size of all the minimal adjustment sets found by $CEELS$ is 2 whereas the number of covariates used by PRE is 24, and the sizes of $\hat{\mathbf{X}}_{\rightarrow W}$, $\hat{Q}_{\rightarrow W}$, $\hat{\mathbf{X}}_{\rightarrow Y}$, $\hat{Z}_{\rightarrow Y}$, $\hat{\mathbf{X}}_{\rightarrow W, Y}$ are 7, 4, 8, 3, 15, respectively. The sizes of the adjustment sets found by EHS are 2 to 6, and 0 to 3 by LVIDA-rFCI.

\begin{table}[t]
\centering
\footnotesize
\caption{The estimated $ACE$s and RE with the IHDP and Twins datasets. The lowest RE in each dataset is boldfaced. $CEELS$ outperforms all comparisons on both datasets.}
\begin{tabular}{lcc||cc}
	\hline
	\multirow{2}{*}{Methods}&
	\multicolumn{2}{c||}{IHDP}&\multicolumn{2}{c}{Twins}\cr\cline{2-5}  &  $ACE$  &  RE (\%) &      $ACE$    &  RE (\%) \\ \hline
	$CEELS$ &   \textbf{4.51} &  \textbf{3.49}\% &  \textbf{-0.02348}  &  \textbf{5.66}\%     \\
	PRE               & 3.94  &  9.61\%                 &   -0.01597  & 35.83\%         \\
	$\hat{\mathbf{X}}_{\rightarrow W}$   &  3.59 &  17.71\%  &   -0.02123  & 14.70\%     \\
	$\hat{\mathbf{Q}}_{\rightarrow W}$            &  3.69 &  15.19\%  &   -0.01946  & 21.84\%    \\
	$\hat{\mathbf{X}}_{\rightarrow Y}$   &  3.71 &  14.96\%  &   -0.01326  & 46.72\%    \\
	$\hat{\mathbf{Z}}_{\rightarrow Y}$            &  3.70 &  15.18\%  &   -0.01228  & 50.65\%    \\
	$\hat{\mathbf{X}}_{\rightarrow W, Y}$&  4.13 &  5.26\%   &   -0.01702  & 31.63\%     \\
	EHS                                 &  4.05 &  7.07\%   &   -0.03287  & 24.27\%     \\
	LVIDA-rFCI                               &  3.81 &  14.58\%  &   -0.02127  & 17.03\%      \\ \hline
\end{tabular}
\label{tab:IHDP_twins}
\end{table}

\subsection{Time Efficiency}
\label{Subsec:time}
We compare the time efficiency of $CEELS$ with the methods which deal with latent variables, including EHS, LVIDA-FCI and LVIDA-rFCI. All experiments are performed on a PC with 2.6 GHz Intel Core i7 processor and 16 GB of memory (RAM).

\begin{figure}[htbp]
\centering
\begin{minipage}{1\linewidth}\centering
	\includegraphics[scale=0.28]{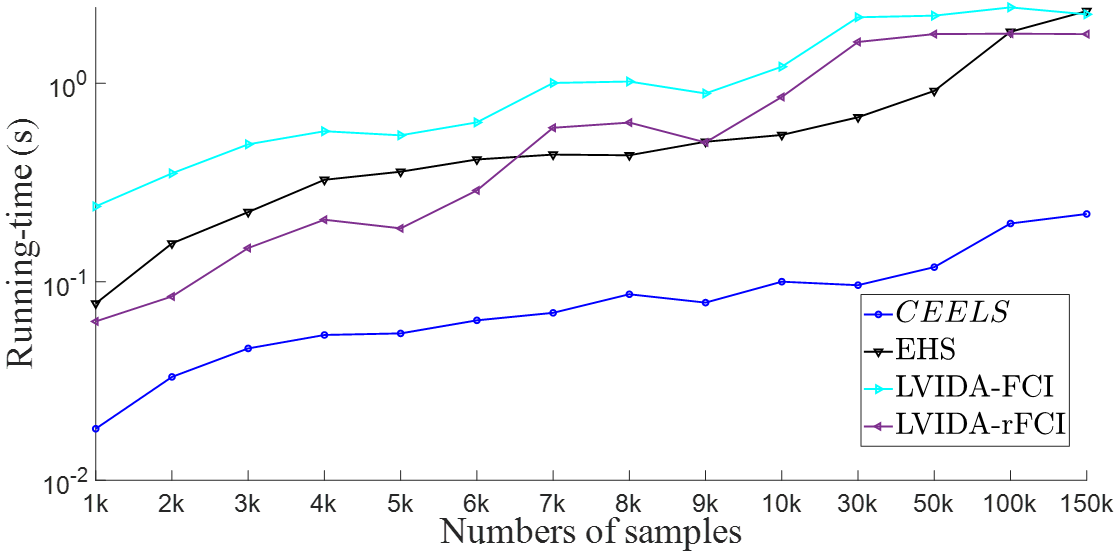}
\end{minipage}
\vfil
\begin{minipage}{1\linewidth}\centering
	\includegraphics[scale=0.28]{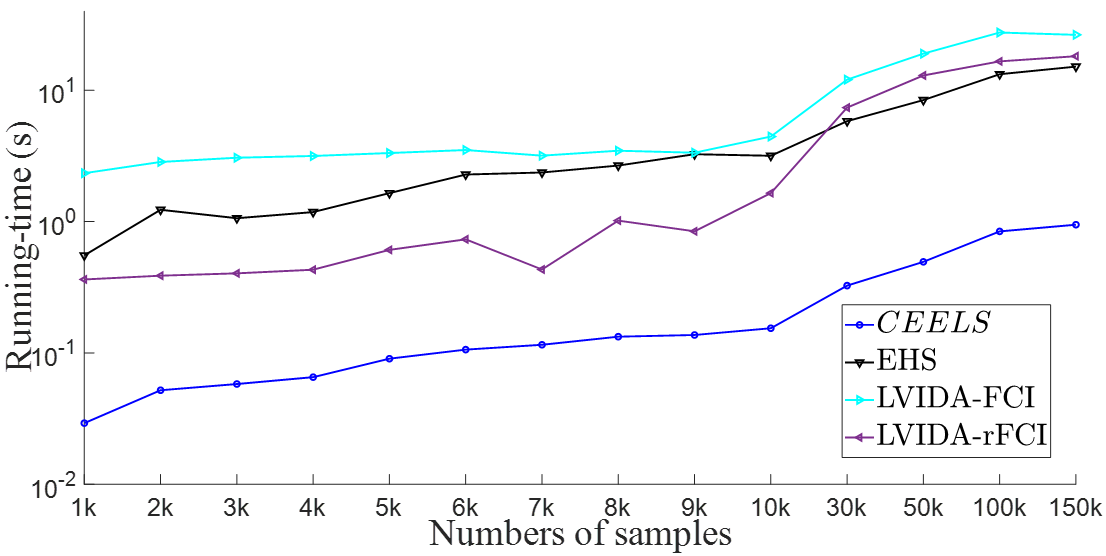}
\end{minipage}
\caption{Runtime in seconds ($\log_{10}$ on the $Y$-axis) of $CEELS$, EHS, LVIDA-FCI and LVIDA-rFCI with the two groups of datasets (top: Group I datasets; bottom: Group II datasets). $CEELS$ is faster than the others on all datasets.}
\label{Result:runningtime}
\end{figure}

We record the running time of $CEELS$, EHS, LVIDA-FCI and LVIDA-rFCI from the experiments on the two groups of synthetic datasets as described in Section~\ref{app:subsec:syndata}) and draw the mean running time in Fig.~\ref{Result:runningtime}. We can see that $CEELS$ is faster than the others. As mentioned in Section~\ref{subsec:bayesiannetwork}, EHS, LVIDA-FCI and LVIDA-rFCI did not return results within 2 hours, but $CEELS$ completed in 5 minutes. Furthermore, the current global structure learning method has a very high time complexity even if some optimised search strategies are used~\cite{chickering2004large,aliferis2010local,cheng2020causal}. Hence, the local search method developed in this paper makes it possible to estimate causal effects unbiasedly from data with latent variables generated by a complex underlying causal graph.

\subsection{The Completeness of $CEELS$}
\label{subsec:completeness}

As mentioned previously, using Corollary~\ref{coro:corCIT:coso} may lead to less complete discovery than using the EHS condition since the local search in Corollary~\ref{coro:corCIT:coso} may miss adjustment sets that satisfy the EHS condition but are not included in the local structure found. In this section, we conduct a simulation to examine the completeness of Corollary~\ref{coro:corCIT:coso} and $CEELS$. The $\mathbf{R}$ packages \emph{pcalg}~\cite{kalisch2012causal} and \emph{ggm}~\cite{sadeghi2012graphical} are utilised to randomly generate MAGs. We generate a DAG and then construct a MAG by removing some common causes in the generated DAG. The edge in the DAG becomes bi-directed in the new graph after removing a common cause. However, the new graph may not be an ancestral graph since there may exist an almost directed cycle. Hence, we repeat the procedure to obtain an ancestral graph. The detailed procedure is described below.

Firstly, we generate a random DAG $\mathcal{G}$ via the function \emph{randomDAG} in the package \emph{pcalg}. There are two parameters, $n$, denoting the number of nodes ($n\geq2$) in the $\mathcal{G}$ and $prob$, representing the probability of connecting a node to another node with higher topological ordering in the $\mathcal{G}$. In our experiments, $n$  is set to 20, 30, 40, and 50, respectively. Theoretically, the average degree of the generated DAG ($d$ for short) is $(n-1)*prob$. In our experiment, $d$ is set to 2, 4, 6, and 8, respectively. When generating DAG $\mathcal{G}$, we limit $W$ to own only one child node $Y$, \ie $W\rightarrow Y$ and other nodes are either the ancestors of $W$, or the ancestors of $Y$, or both.

Secondly, we randomly remove the required number of common causes as latent variables from the generated DAG $\mathcal{G}$. The number of latent variables is denoted as $l$. $l$ is set to 2, 4, 6, 8, 10, and 12, respectively. In this step, we only remove a node with more than two children, \ie common cause. After a node $X$ is removed from the generated DAG $\mathcal{G}$, for each pair in $Ch(X)$, add a bi-directed edge in $\mathcal{G}$, and for each node in $Pa(X)$, add a directed edge to every node in $Ch(X)$. In this way, we obtain a new graph $\mathcal{G}'$. The function \emph{isAG} in the $\mathbf{R}$ package \emph{ggm} is employed to examine whether $\mathcal{G}'$ is an ancestor graph or not. If $\mathcal{G}'$ is an ancestral graph, then the function \emph{Max} in the $\mathbf{R}$ package \emph{ggm} is used to construct a MAG $\mathcal{M}$, otherwise discard $\mathcal{G}'$ and repeat the above procedure.

Finally, we generate a synthetic dataset that is faithful to the constructed MAG $\mathcal{M}$. In this step, the original generated DAG $\mathcal{G}$ is used to generate a synthetic dataset. The function \emph{rmvDAG} in the package \emph{pcalg} is utilised to generate multivariate data with dependency structure specified by the DAG $\mathcal{G}$. The sample size is set to 10,000 for all generated datasets. Then, the variables deleted from the DAG $\mathcal{G}$ to obtain MAG $\mathcal{M}$ are removed from the generated dataset. The generated dataset is for MAG $\mathcal{M}$.

\begin{table}[t]
\centering
\footnotesize
\caption{Experiment results on the completeness of adjustment set discovery using Corollary~\ref{coro:corCIT:coso} and $CEELS$. In the table, the expression A/B (e.g. 99/99) represents the number of trials (out of 100) when an adjustment set is identified using Corollary~\ref{coro:corCIT:coso} (A) and $CEELS$ (B) respectively. $n$ is the number of nodes in a MAG, $l$ indicates the number of latent variables, and $d$ denotes the average degree of the generated DAGs that are used to construct MAGs. `-' denotes that the density of the constructed MAG is more than 0.5.}
\begin{tabular}{|c|c|c|c|c|c|c|c|}
	\hline
	$n$ & $l$                & $d=2$            &     $d=4$       &      $d=6$       &       $d=8$     \\ \hline
	\multirow{5}{*}{20} & 2  & 99 / 99 & 98 / 100&  93 / 97 & 92 / 98 \\ \cline{2-6}
	& 4  & 100 / 100 & 97 / 97  & 96 / 98 & -  \\ \cline{2-6}
	& 6  & 98 / 98   & 97 / 97  & 94 / 95 & -  \\ \cline{2-6}
	& 8  & 99 / 99   & 95 / 96  & -       & - \\ \cline{2-6}
	& 10 & 97 / 97   & 95 / 96  & -       & - \\ \hline
	\multirow{6}{*}{30} &  2 & 100 / 100 & 100 / 100& 96 / 98 & 95 / 97   \\ \cline{2-6}
	& 4  & 100 / 100 & 100 / 100& 98 / 99 & 95 / 98    \\ \cline{2-6}
	& 6  & 100 / 100 & 99 / 99  & 99 / 100& 93 / 93  \\ \cline{2-6}
	& 8  & 99 / 99   & 100 / 100& 96 / 98 & 94 / 97\\ \cline{2-6}
	& 10 & 100 / 100 & 99 / 100 & 97 / 98 & 95 / 96 \\ \cline{2-6}
	& 12 & 99 / 99   & 97 / 97  & 98 / 99 & 93 / 93 \\ \hline
	\multirow{7}{*}{40} & 2  & 99 / 99   & 100 / 100& 99 / 100 &98 / 100 \\ \cline{2-6}
	& 4  & 100 / 100 &100 / 100 &98 / 100  &96 / 97 \\ \cline{2-6}
	& 6  & 100 / 100 &99 / 99   &100 / 100 &96 / 98 \\ \cline{2-6}
	& 8  & 100 / 100 &100 / 100 &99 / 100  &96 / 97  \\ \cline{2-6}
	& 10 & 100 / 100 & 99 / 99  &100 / 100 &95 / 99 \\ \cline{2-6}
	& 12 & 98 / 98  & 100 / 100 &98 / 99   & 95 / 96 \\ \hline
	\multirow{8}{*}{50} & 2  & 100 / 100 & 100 / 100& 100 / 100 & 99 / 99 \\ \cline{2-6}
	& 4  & 100 / 100 & 100 / 100& 99 / 100 & 98 / 99 \\ \cline{2-6}
	& 6  & 100 / 100 & 100 / 100& 99 / 100  & 96 / 97    \\ \cline{2-6}
	& 8  & 100 / 100 & 100 / 100& 99 / 99   & 98 / 100 \\ \cline{2-6}
	& 10 & 99 / 99   & 100 / 100& 99 / 100 & 98 / 99   \\ \cline{2-6}
	& 12 & 100 / 100 & 100 / 100& 99 / 99   & 97 / 98 \\ \hline
\end{tabular}
\label{tab:completeness}
\end{table}
For each setting (number of nodes, average node degree and number of deleted common causes), we randomly construct 100 MAGs and generate the corresponding synthetic datasets. We run Corollary~\ref{coro:corCIT:coso} (which is implemented based on $CEELS$ by replacing Line 2 of Algorithm~\ref{pseudocode01} with $Adj'(W)=Adj(W)\setminus\{Y\}$) and $CEELS$ algorithm on the datasets to count the number of cases in which an adjustment set is identified. The experimental results are reported in Table~\ref{tab:completeness}. In a setting, if the constructed MAG is too dense (the density is more than 0.5), we do not report its results (indicated by a dash symbol in Table~\ref{tab:completeness}).

From Table~\ref{tab:completeness}, we observe that in most cases, adjustment sets are found by local search using Corollary~\ref{coro:corCIT:coso}. Since $CEELS$ searches in the expanded local space, there are more cases when $CEELS$ can find adjustment sets. Overall, the completeness rates of both Corollary~\ref{coro:corCIT:coso} and $CEELS$ are close to 100\% in all experimental situations.

\section{Discussion}
The proposed $CEELS$ algorithm assumes that the Markov property and faithfulness both hold. However, it is not clear what would be the implication if the assumptions are violated in practice because they are not testable in data~\cite{pearl2009causality}. The Markov property and faithfulness are beliefs that we need to have to use graphical causal methods. Consequently, we cannot test the validity of the results of the causal inference using data alone as we normally do in evaluating the accuracies of prediction/classification models.

The best way to test the validity of the results of causal inference is via RCTs. When an RCT is not possible, a check of consistency with existing domain knowledge is an alternative. The consistency of a result can be checked using some independent datasets obtained in different environments, such as datasets from other labs for a biological study on the same topic. A causal result is supposed to be invariant across the multiple independent  datasets.

When a dataset does not represent a population, if we use the data to infer a causal graph or estimate a causal effect, we will get a biased result. This bias is again not testable in data. The above suggested means are useful to test the validity of a result. When using data to infer causal graphs, the failures of independence tests account for most errors. The errors can be reduced by increasing the sample size. Our simulation experiment also demonstrates such error reduction.

Additionally, $CEELS$ requires a COSO variable. If there is not a COSO variable in the data, $CEELS$ will fail. We have argued that a COSO variable is likely to exist in many real-world applications, but this could still be a restriction. For example, a COSO variable given by domain experts will help with the accurate estimation of causal effect, but a wrongly specified COSO variable will lead to biased results.

Practically, in many areas of economics, epidemiology, health care and social studies, domain experts often know one or more direct causes of the treatment variable (and know that they are not direct causes of the outcome), i.e. domain experts are able to nominate COSO variables. Our algorithm can then utilise the domain knowledge and the data to identify precise causal effects in data efficiently. This gives domain experts an effective tool to estimate causal effect in data without worrying about other latent variables and potentially saves them a lot of unnecessary costly trials and experiments.

\section{Related Work}
\label{sec:relatedwork}
Estimating the causal effect of an intervention (or treatment) on an outcome of interest is a major task of causal inference~\cite{witte2019covariate}. A causal effect can be an Average Causal Effect (ACE) in the whole population or a Conditional Average Causal Effect (CACE) in a subpopulation. We only consider the former in this section since the focus of this paper is on ACE. Confounding adjustment is a common way of obtaining unbiased estimation, and we will discuss the methods around confounding adjustment, which is the approach taken by our work. There are two main frameworks for estimating causal effects: potential outcome framework~\cite{rubin1974estimating,rosenbaum1983central,imbens2015causal}, graphical causal modelling~\cite{spirtes2000causation,pearl2009causality,koller2009probabilistic}. 

When an adjustment set is given and the unconfoundedness (\aka ``strong ignorability'') is assumed~\cite{rubin1974estimating,rosenbaum1983central}, there are many methods based on the potential outcome framework~\cite{rubin1974estimating,rosenbaum1983central,hill2011bayesian,imbens2015causal,wager2018estimation,kuang2020data}. Rosenbaum and Rubin~\cite{rosenbaum1983central} proposed the propensity score, which is commonly used for confounding control for estimating causal effects. Then, many Machine Learning based methods are utilised to calculate the propensity score~\cite{bloniarz2016lasso,lee2010improving}. Bang and Robins proposed a doubly robust estimator by combining the propensity weighting and regression~\cite{bang2005doubly}. A comprehensive review can be found in~\cite{guo2020survey}. These methods do not deal with latent variables, and they may fail when the data contains $M$-bias~\cite{greenland2003quantifying,pearl2009myth,sjolander2009propensity}. Our approach differs from these methods because it deals with latent variables and handles $M$-bias as shown in our experiments. 

When an adjustment set is not given, and the causal graphs (DAG or MAG) for the data generation mechanisms are given, the back-door criterion for DAG~\cite{pearl1995causal}, adjustment criterion for DAG~\cite{shpitser2012validity}, generalised back-door criterion for MAG~\cite{maathuis2015generalized} and generalised adjustment criterion for MAG~\cite{perkovic2018complete} can be used to determine a proper adjustment set.

However, the complete causal graph is rarely known in real-world applications. Hence, data-driven algorithms for finding adjustment sets are necessary, and we have reviewed the related data-driven algorithms based on graphical causal modelling~\cite{maathuis2009estimating,hyttinen2015calculus,entner2013data,malinsky2017estimating,cheng2020causal} to obtain unbiased causal effects directly from data in the Introduction section, so we skip them here. 

Some methods based on the LiNGAM models~\cite{shimizu2006linear,shimizu2011directlingam,tashiro2014parcelingam,chen2021causal}  are data-driven and handle latent variable. These methods recover the causal graph and estimate some parameters of causal strengths simultaneously under the assumption that the causal graphs/models are linear, acyclic, and with non-Gaussian error terms. One advantage of these methods is that they are able to provide a unique causal DAG/MAG or a small set of Markov equivalent DAGs/MAGs. However, the assumptions are strong, and the methods may not be feasible for all applications. Furthermore, these methods have a very high computational complexity and they are infeasible in many applications.

\section{Conclusion}
\label{sec:con}
In this paper, we have identified a practical problem setting where valid adjustment sets can be found via local search in data with latent variables, which enables unique and unbiased causal effect estimation in a fast way. We have developed the theorems and corollaries to ensure the soundness of the local search. Based on the developed theorems and corollaries, we have designed an efficient algorithm, $CEELS$, for finding adjustment sets and estimating causal effects from data with latent variables. A level-wise search strategy is employed to further improve efficiency and to find minimal adjustment sets via local causal structure discovery and search. We have conducted experiments on synthetic and real-world datasets to evaluate the performance of $CEELS$. The results have demonstrated that $CEELS$ can obtain a more accurate causal effect estimation than the other existing methods, and $CEELS$ is much faster than other methods that deal with latent variables.

\bibliographystyle{IEEEtran}
\bibliography{CEELS.bib}
\end{document}